\documentclass[lettersize,journal]{IEEEtran}
\usepackage{amsmath,amsfonts}
\usepackage{algorithmic}
\usepackage{algorithm}
\usepackage{array}
\usepackage[caption=false,font=normalsize,labelfont=sf,textfont=sf]{subfig}
\usepackage{textcomp}
\usepackage{stfloats}
\usepackage{url}
\usepackage{verbatim}
\usepackage{graphicx}
\usepackage{color}
\usepackage{scalefnt}
\usepackage{cite}
\begin{document} 
\scalefont{0.985}
\title{\huge{On the Achievable Error Rate Performance of Pilot-Aided Simultaneous Communication and Localisation} }

\author{Shuaishuai Han,~\IEEEmembership{Student Member,~IEEE,} Emad Alsusa, 
\IEEEmembership{Senior Member,~IEEE,} Mohammad Ahmad Al-Jarrah,~\IEEEmembership{Member,~IEEE,}  Mahmoud AlaaEldin,~\IEEEmembership{Member,~IEEE,}

\thanks
{Shuaishuai Han, E. Alsusa, M. A. Al-Jarrah and Mahmoud AlaaEldin are with the Department of Electrical and Electronic Engineering,
University of Manchester, Manchester M13 9PL, U.K. (e-mail: {shuaishuai.han@postgrad.manchester.ac.uk,}
\{e.alsusa, mohammad.al-jarrah,  mahmoud.alaaeldin\}@manchester.ac.uk).}\thanks}

\date{}
\maketitle

\begin{abstract}
\textbf
This paper investigates the symbol error rate (SER) performance of the pilot-aided simultaneous communication and localisation (PASCAL) system. A scenario where multiple drones transmit communication signals to a base station (BS), which needs to simultaneously decode the signals and continuously locate the drones' positions during the communication session, is considered. The BS operates in two stages: first, it estimates the drones' location parameters using pilot signals; second, it performs data detection by reconstructing the channel response based on the estimated location parameters. The theoretical analysis presented demonstrates that the estimated location parameters follow Gaussian distributions with means equal to the actual values and variances determined by the root mean square error (RMSE) of the estimator. Using these distributions, the average SER is derived to quantify the impact of localisation errors on decoding performance. This analysis highlights the synergy between communication and localisation, providing valuable insights into the influence of localisation inaccuracies on the performance of location-aware communication systems. Simulations are conducted to validate the theoretical derivations.
\end{abstract}

\section{Introduction}
With the global commercial deployment of fifth-generation (5G) mobile communication systems, research into technologies beyond 5G has been progressing rapidly. Among the emerging technologies, integrated sensing and communication (ISAC) stands out as a key enabler for next-generation wireless networks. ISAC integrates sensing and communication functions on a shared platform, utilizing the same network resources and signal processing modules \cite{ISAC_Wang}. Compared to traditionally separated sensing and communication systems, ISAC offers significant advantages in spectrum, energy, and hardware efficiency. However, much of the current research on ISAC overlooks the fact that many localisation targets, such as drones and vehicles, are equipped with their own transceivers. These targets can actively transmit signals to the base station (BS) not only for communication but also to facilitate localisation using pilot signals. Building on this concept, we propose the pilot-aided simultaneous communication and localisation (PASCAL) system. Compared to the existing ISAC systems \cite{ISAC_an, ISAC_hua,ISACJarrah1}, PASCAL is more energy efficient as it only entails one-way path-loss, while ISAC typically transmit signals to the targets and rely on the echos for localisation, resulting in higher energy consumption.

Since the advent of radar systems, localisation technology has advanced significantly over the past few decades. Today, localisation can be achieved through various methods, including global positioning system (GPS)-based techniques, vision-based techniques, radar-based techniques, and more \cite{localisation_Lu}. However, the performance of GPS-based localisation may deteriorate significantly in the GPS-denied environment such as indoor or underground due to signal blocking and serious multipath fading \cite{GPS}. Compared to GPS-based localisation, vision-based localisation estimates the location parameters of targets using images from monocular or binocular cameras and can be deployed in a GPS-denied environment. The rich environmental information within images can provide high-precision localisation performance under well-lit conditions, but it requires considerable memory and computational resources. In this paper, we focus on radar-based localisation by continuously estimating the location of targets using radars during the movement of targets, which is not constrained by GPS-denial and is more efficient in terms of memory and computational resources than vision-based techniques \cite{localisation_Lu}.

The advancement of wireless communication systems has significantly promoted global connectivity and social development. However, achieving high-quality communication performance heavily depends on reliable channel estimation. Channel estimation methods are typically categorized as parametric or non-parametric, based on the assumptions underlying the channel model. Parametric channel estimation models the channel using a set of parameters, such as path gains and path delays, which are then used to reconstruct the channel response. In contrast, non-parametric channel estimation does not rely on such a model; instead, it directly estimates the channel frequency response. \cite{channelestimation_wei}. Non-parametric channel estimation performs better than parametric channel estimation in complex and rapidly changing channels. However, parametric channel estimation has been proven in \cite{parametric channel estimation1, parametric channel estimation} to provide more accurate channel information for sparse channels by employing fewer pilot signals than non-parametric estimation. It has been shown in \cite{parametric channel estimation2, parametric channel estimation3} that the mean square error (MSE) of channel estimators using the parametric method is superior to that of non-parametric approaches. Hence, in this paper, we adopt the parametric channel estimation method, leveraging estimated location parameters to infer channel information. This approach recognizes that location information constitutes a significant component of channel state information (CSI), particularly in line-of-sight (LoS) scenarios. Moreover, parametric channel estimation is especially well-suited for the PASCAL system, as the system inherently requires location estimation which can be directly utilised to obtain CSI.

\subsection{Related Works}
In recent years, there has been a surge of interest in ISAC techniques due to their significant advantages, such as shared spectrum and equipment, compared to separate designs. ISAC not only improves efficiency but also ensures non-interference between communication and sensing systems. However, despite numerous studies implementing communication and sensing functionalities on the same platform, these functions are often applied to separate objects, limiting their ability to establish a collaborative relationship. For instance, the authors of \cite{ISAC_ouyang2, ISAC_yuan} utilize the receiver array of a radar-communications BS to simultaneously receive echoes for sensing target locations and signals from users to facilitate uplink communication. Even when both communication and sensing are achieved by the same receive array, they are utilised to serve different objects. A similar case is considered in \cite{ISAC_wang3, ISAC_Lyu, ISAC_meng}, where an ISAC BS simultaneously performs downlink communication with users and localizes other objects by transmitting sensing signals and receiving echoes. In \cite{ISAC_wang2}, even though the authors employ the BS to transmit ISAC signals to communicate with multiple vehicles and also receive echoes to track the locations of the same vehicles, they still ignore the possible cooperation between the communication and localisation functions. As a consequence, the estimated location information by the aforementioned research has not been efficiently utilised. Additionally, in such systems \cite{ISAC_ouyang2, ISAC_yuan, ISAC_wang3, ISAC_Lyu, ISAC_meng, ISAC_wang2}, communication and localisation functionalities coexist as two separate services without achieving a comprehensive integration in signal utilization, which is a key focus of the PASCAL system.

Location information is typically obtained by receiving radar waveforms (e.g., chirp signals and frequency/phase-coded waveforms) reflected from the targets \cite{ISAC_liu}. However, recent studies in ISAC have increasingly focused on using built-in pilots within communication signals to acquire location information. In \cite{localisation_gao}, the pilot signal is embedded into the data frame for estimating the Doppler frequency information. In addition to obtaining the Doppler frequency, the authors in \cite{localisation_zhang} exploited the pilots to also estimate the delay spread, which are then used to indicate the distance and velocity of the target, respectively. In \cite{localisation_huang}, pilot signals are transmitted from the BS towards the targets to obtain an initial estimate of the locations of the targets. In \cite{localisation_bao, localisation_Ozkaptan}, the authors employ pilots to estimate the channel and localize targets simultaneously. Likewise, in \cite{localisation_zhu}, the pilot signals are adopted for ranging and channel estimation. The authors in \cite{localisation_gao}-\cite{localisation_zhu} also utilize the pilot signals to estimate various location parameters. Nonetheless, the aforementioned references \cite{localisation_gao}-\cite{localisation_zhu} rely on echos from the pilot signals to achieve localisation of targets. As a consequence, their approach for acquiring the location information suffers from round-trip path-loss, hence it is less efficient compared to the PASCAL system which utilizes the targets' transmitted signals to achieve localisation.

\subsection{Motivation and Contributions}
Recently, in \cite{ISAC_Han}, we introduced a PASCAL system where multiple drones actively transmit signals to the BS, which processes the received signals to obtain the location information of the drones from the pilots, and also decodes the symbols contained within the signals. Compared to the conventional ISAC systems in \cite{ISAC_ouyang2}-\cite{localisation_zhu} which utilize the reflected signals for localisation, PASCAL is more energy-efficient as it suffers from one-way path-loss rather than round-way path-loss. Motivated by the efficiency of the PASCAL system, this paper presents a symbol error rate (SER) analysis to assess the communication reliability and accuracy of the system. The SER analysis is crucial because the location information obtained is used to infer channel state information, making decoding performance highly dependent on location accuracy. By conducting the SER analysis, we not only gain insight into the synergistic relationship between communication and localisation within the PASCAL system but also evaluate how estimation errors in location parameters impact communication reliability. Additionally, this analysis is valuable for understanding the effect of location errors on communication performance in systems that heavily rely on location accuracy, such as location-aware services \cite{location aware services}. Compared to our work in \cite{ISAC_Han}, which focuses on evaluating the effect of gain-phase defects on the localisation accuracy and the achievable sum data rate of the PASCAL system, and other related literature, the main contributions of this paper can be summarized as follows.

\begin{enumerate}
\item We present a mathematical analysis of SER to evaluate the impact of estimation errors in various location parameters (i.e., angle, Doppler frequency, and range) on the data decoding performance of the PASCAL system. 

\item We also analyze the statistical characteristics of the estimated location parameters, including angle, Doppler frequency, and range, under Gaussian noise. Our analysis shows that the estimated location parameters follow independent Gaussian distributions, with means equal to their true values and variances corresponding to the root mean square error (RMSE). These statistical characteristics are then used to derive the average symbol error rate (SER), which is shown to be a function of the RMSE. 

\item As the estimated location parameters follow different Gaussian distributions with distinct means and variances, determining the overall distribution of the channel constructed with multiple location parameters becomes challenging, increasing the complexity of the analysis. To address this, we first derive a conditional SER based on the estimation errors of the location parameters. This is then evaluated over the error distributions, and SER is obtained using a Taylor approximation with fast convergence.  

\item Analytical and simulated results are presented to demonstrate the impact of localisation errors on communication accuracy. The results show that improved localisation accuracy enhances the average SER. Additionally, the influence of the number of pilots on the PASCAL system is examined. The findings indicate that an increased number of pilots benefits both SER and localisation accuracy. Finally, the excellent agreement between the analytical and simulation results confirms the accuracy of our analysis.
\color{black}
\end{enumerate}

The rest of the paper is organised as follows. In Sec. \ref{system model}, the system model for the PASCAL system is provided. Sec. \ref{localisation and communication} illustrates the ML-based localisation algorithm and data detection with maximum ratio combining (MRC). Sec. \ref{performance analysis DTDD} presents the performance analysis for evaluating the localisation and communication performance of the PASCAL system. Sec. \ref{results} demonstrates the simulation and analytical results, and Sec. \ref{Conclusion} concludes this paper.

$Notations$: $\hat{\boldsymbol{ \psi}}$ denotes a vector composed of estimated parameters. $\left \| \cdot \right \|_2$ represents the Euclidean norm. $\mathbb{E}[\cdot]$ refers to the statistical expectation. $\det(\cdot)$ represents the determinant of a matrix. ${[ \cdot ]^*}$, ${[ \cdot ]^T}$ and ${[ \cdot ]^H}$ indicate the copmplex conjugate, transposition and Hermitian transposition. 

\begin{figure}[ptb]
	\centering
	\includegraphics [scale=0.41]{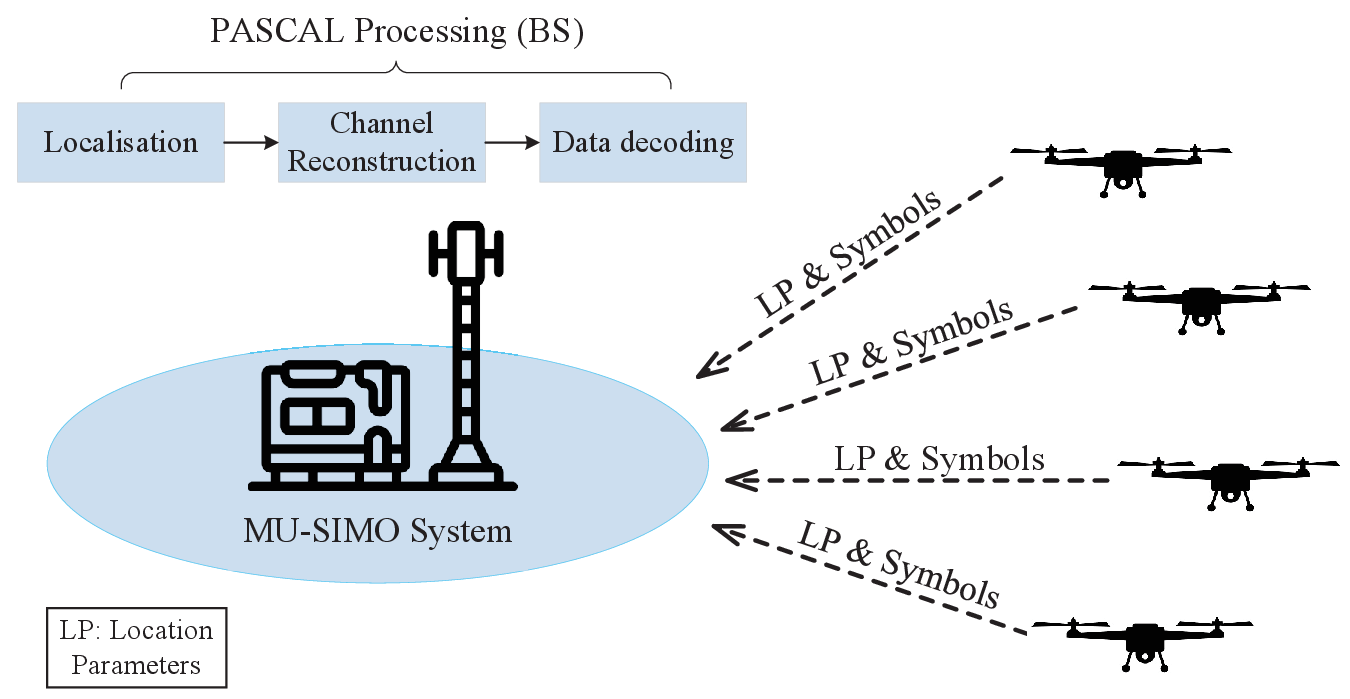}
\caption{System model of the PASCAL system.}%
	\label{system_model}
\end{figure}

\section{System Model}
\label{system model}
As shown in Fig. \ref{system_model}, in PASCAL, we consider a multiuser single-input multiple-output (MU-SIMO) system consisting of $K$ moving single-antenna drones located in the far field which send information signals to a BS with multiple antennas, where the latter aims to localize the drones using the pilot signals and decode information symbols. The above system model is applicable to scenarios such as GPS-denied environments \cite{GPS-denied} including tunnel, forest and underground environments where the precise GPS information of the drones is not available. Furthermore, commercial GPS may not be accurate in localizing drones. For instance, the estimation errors for the altitude of the drones could reach 15 m. Such
amount of error is considerable especially when the density
of deployed drones is high and could lead to drones' clashes
under these circumstances. As a consequence, the drones in the PASCAL system actively send information signals to the BS which leverages these signals to facilitate the localisation of the drones. 
Given the high likelihood of Line-of-Sight (LoS) channel availability in air-to-ground (A2G) links within drone-based ISAC systems \cite{ISAC_Lyu}, and the potential impact of undesired Non-LoS (nLoS) paths on reliable target identification in MIMO radar-based localisation—commonly referred to as the virtual or ghost targets phenomenon \cite{ghost}— we utilize ray tracing \cite{ray tracing} to pre-process the received signal. This approach extracts the LoS component from the multipath signals prior to localisation, improving accuracy.

\begin{figure}[ptb]
	\centering
	\includegraphics [scale=0.42]{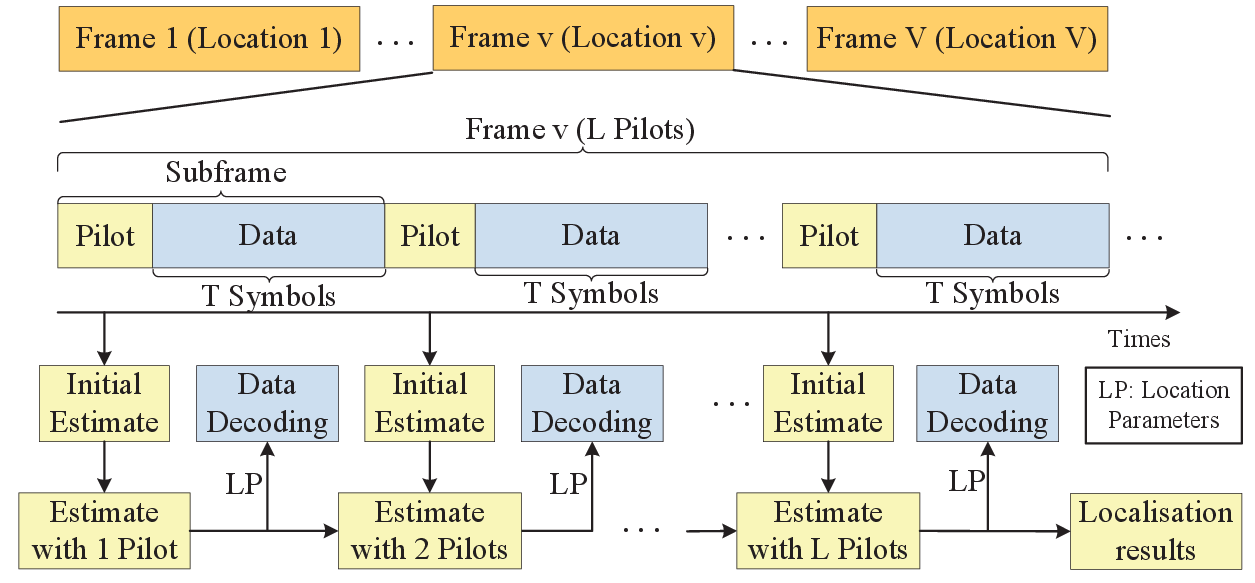}
\caption{Frame structure for the PASCAL system.}%
	\label{PASCAL}

\end{figure}

In Fig. \ref{PASCAL}, the frame structure of the PASCAL system is designed to achieve efficient localisation and data decoding, comprising a total of $V$ frames. It is assumed that the velocities and locations of the drones remain constant within each frame, an assumption consistent with those used in prior works on similar systems \cite{localisation_gao}, while the drones are located at different positions in different frames. This assumption is reasonable as, for instance, a frame with a length of $1000$ symbols only takes up $1\times 10^{-3}$ seconds when the symbol rate is $10^6$ symbols per second. In addition, each frame consists of $L$ subframes and each subframe contains one pilot and $T$ symbols. In addition, the pilot signals embedded within the data symbols to estimates of the location parameters. Specifically, the first pilot is employed to estimate the location parameters, which are then applied in decoding the symbols in the first subframe. Afterwards, both the first and the second pilots are usedto estimate the locations, which are subsequently utilised to decode the symbols in the second subframe. In the $l$th subframe, $l$ pilots are utilised to estimate the location parameters, which are then employed for decoding the $T$ symbols in that subframe. In the last subframe, the final estimation of location parameters can be obtained, which is then utilised to decode the symbols in the last subframe.

Once the information signals are received by the BS, it will decode the symbols and extract the location information including direction of arrival (DOA), range and Doppler frequency. The DOA, range and Doppler frequency of drone $k$ can be denoted by using $\theta_k$, $d_k$ and $f_{D,k}$.  It is worth noting that the Doppler frequency is employed in this paper to model the impact of the velocity of drones on the received signals. Since the velocities and locations of the drones may remain constant for a short period, $\theta_k$, $d_k$ and $f_{D,k}$ for $k\in\{1,..,K\}$ can be assumed to be time-invariant during each frame.  In addition, the BS used in this paper is composed of a uniform linear array (ULA) with ${N}$ antennas, where the adjacent antennas are separated with a distance of half of the wavelength, i.e., $d_0=\lambda /2$.     

By arranging the outputs of the matched filter, the received signal for the $t$th transmission within the $l$th subframe from all drones, where $t\in\{0,...,T\}$ and $l\in\{1,...,L\}$,  is denoted by
\begin{equation} 
	\label{received signal matrix2}
	{\bf{  y}}_{t,l} = {\bf{A}}{{\boldsymbol{\omega}}(l)}  {\bf{s}}_{t,l}+ {\bf{  n}}, 
\end{equation}
where ${\bf{  y}}_{t,l} \in \mathbb{C}{^{N \times 1}}$ contains the signals received by the $N$ antennas of the BS and ${\bf{A}}{{\boldsymbol{\omega}}(l)}\in \mathbb{C}{^{N \times K}}$ indicates the LoS channel response. ${\bf{s}}_{t,l}\in \mathbb{C}{^{K \times 1}}$ and ${{\bf{n}}}\in \mathbb{C}{^{N \times 1}}$ denote the vectors with information signals (i.e., pilot signal and symbol signals) and additive white Gaussian noise (AWGN), respectively. ${\bf{  s}}_{t,l}$ is given by ${\bf{  s}}_{t,l} = [\sqrt{{{P}}_{1}}{s_{t,l,1}},...,\sqrt{{{P}}_{K}}{s_{t,l,K}}]^T$, where ${{{P}}_{k}}$ represents the transmit power of drone $k$ and ${s_{t,l,k}}$ refers to the $t$th signal in the $l$th subframe from drone $k$. When $t=0$,  ${\bf{s}}_{0,l}$ denotes the vector of pilot signal in the $l$th subframe. Since ${{{s}}_{0,l,k}}=1$ for $k\in\{1,...,k\}$, ${{\bf{s}}_{0,l}}=[ \sqrt{{{P}}_{1}},...,\sqrt{{{P}}_{K}}  ]^T$. ${\boldsymbol{\omega}}(l)$ refers to a diagonal matrix with path loss and Doppler frequency, which is written as 
\begin{equation}
   {\boldsymbol{\omega}}(l) \stackrel{\triangle}{=} {\mathrm{diag}\{{\eta _1}{{\mathrm{e}}^{j2\pi {f_{D,1}}l/{f_s}}},...{\eta _K}{{\mathrm{e}}^{j2\pi {f_{D,K}}l/{f_s}}}\}}, 
\end{equation}
where ${f_s}$ denotes the signal sampling frequency and ${\eta _k}$
indicates the free space path loss with the definition ${\eta _k}=\frac{\lambda}{4 \pi d_k}$, $d_k $ represents the distance between drone $k$ and the BS. ${\bf{A}}$ represents the array manifold of the BS, which can be denoted by ${\bf{A}} =\big[{{\bf{a}}}({\theta _1}) ,...,{{\bf{ a}}}({\theta _K}) ]$, in which the steering vector ${{\bf{a}}}({\theta _k})$ can be expressed as
\begin{equation} 
	\begin{array}{l}
		  {{\bf{a}}}({\theta _k}) = [{{{a}}}_{1}({\theta _k}),...,{{{a}}}_{N}({\theta _k})]^T,
	\end{array}
\end{equation}
where $		\displaystyle{{{a}}}_{n}({\theta _k})\overset{\triangle}{=} {{\mathrm{e}}^{- j2\pi  ({n}-1) d_0\sin {\theta _k}/\lambda}}\;\forall n\in\{1,..,N\}$.

\section{Simultaneous Communication And localisation}
\label{localisation and communication}
 
\subsection{Maximum Likelihood based Algorithm}
\label{ML localisation}
In this section, an ML-based algorithm is proposed to achieve drone localisation by continuously estimating location parameters across multiple frames. The algorithm adapts to changes in drone locations and velocities, which occur at each frame cycle, by utilizing the increasing number of pilots with each sub/frame. This approach ensures improved accuracy in estimating dynamic changes in the drones' positions over time. In the $l$th subframe of frame $v$, the pilot signal vector with $l$ pilots can be given by
\vspace{-0.2cm}
\begin{equation}
\label{X1}
{{\bf{ y}}_1} = {\left\{\Big\{{{\bf{A}}}{\boldsymbol{\omega}}(1){\bf{s}}_{0,1}\Big\}^T,...,\Big\{{{\bf{A}}}{\boldsymbol{\omega}}({l}){\bf{s}}_{0,l}\Big\}^T\right\}^T} + {{\bf{ n}}_1}, 
\vspace{-0.2cm}
\end{equation}
where ${{\bf{ y}}_1}\in {{\mathbb{C}}^{{N}{l} \times 1}}$ follows a multivariate Gaussian distribution in the case of Gaussian noise and thus its PDF is shown as
\begin{equation}
\label{y PDF}
f({{\bf{ y}}_1}|\boldsymbol{\psi} ) = \frac{1}{{{{\pi ^{N{l}}}\det ({\bf{\Gamma }})}}}{{e}^{ - \frac{{\left[{{\bf{ y}}_1} - \boldsymbol{\mu} \right]}^H\left[{{\bf{ y}}_1} - \boldsymbol{\mu} \right]}{{\bf{\Gamma }}}}},
\vspace{-0.2cm}
\end{equation}
where $\boldsymbol{\psi} = {[{{\boldsymbol{\theta }}^T},{{\boldsymbol{d }}^T},{{\boldsymbol{f_{D}}}^T}]^T}$ represents a vector composed of the deterministic unknown location parameters, in which $\boldsymbol \theta=[\theta_1,...,\theta_K] $, $\boldsymbol d=[d_1,...,d_K] $ and $ \boldsymbol{f_{D}} =[f_{D,1},...,f_{D,K}]  $. $\boldsymbol{\mu}$ and ${\bf{\Gamma }}$ indicate the mean vector and covariance matrix, respectively. $\boldsymbol{\mu}={\left\{\{{{\bf{A}}}{\boldsymbol{\omega}}(1){\bf{s}}_{0,1}\}^T,...,\{{{\bf{A}}}{\boldsymbol{\omega}}({l}){\bf{s}}_{0,l}\}^T\right\}^T}$ and ${\bf{\Gamma }}=\sigma^2\bf{I}$, in which $\sigma^2$ and $\bf{I}$ refer to the variance of AWGN and the identity matrix.

By using the PDF in \eqref{y PDF}, the maximum likelihood estimator (MLE) can be obtained as
\vspace{-0.2cm}
\begin{equation}
\label{True ML}
\begin{array}{*{20}{l}}
[ \hat {\boldsymbol{\psi}} ] = \mathop {\arg \max }\limits_{{\boldsymbol{\psi}} } \ln f({\bf{ y}}_1|{\boldsymbol{\psi }})
\\ \ \ \ \ \ \ \  
= \mathop {\arg \min }\limits_{{\boldsymbol{\psi}}} ||{\bf{ y}}_1 - {\boldsymbol{\mu }}||_2^2.
\end{array}
\vspace{-0.2cm}
\end{equation}
where the output of MLE in \eqref{True ML} provides the estimation results of the location parameters in the $l$th subframe of frame $v$, which can be denoted by $\hat{ \boldsymbol{\psi}}_{l}$. With the increase in the number of pilots, the localisation performance can be improved, thus
a more accurate estimation result will be utilised for data decoding to improve performance. Nonetheless, the final estimation of the location parameters is obtained by using all the pilots (i.e., $l=L$) in each frame, as shown in Fig. \ref{PASCAL}.

\begin{figure}[ptb]
\centering 

{\includegraphics  [height=1.4in, width=2.736in]{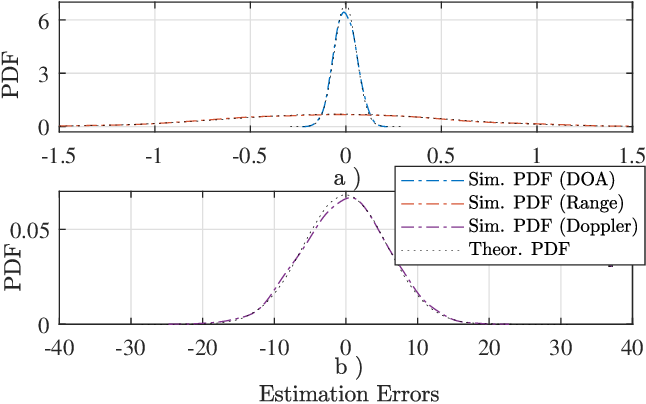}}

 \caption{PDF of the location parameters estimation errors under AWGN noise.} 
\label{PDF}
\vspace{-0.6cm}
\end{figure}

\vspace{-0.2cm}
\subsection{ Data Decoding}
In the communication stage, the estimated location parameters obtained from Sec. \ref{ML localisation} are employed to infer the channel responses in each subframe by using the parametric channel estimation method, which are then fed into an MRC to combine the received signals from receiving antennas and then detect the symbols. According to the MRC principle, in the $l$th subframe, the received signal ${{\bf{ y}}_{t,l}}$ is multiplied by the Hermitian of the estimated channel response matrix ${\bf{\hat H}}$ as ${\bf{ x}}_{t,l} = {\bf{\hat H}}^H{{\bf{ y}}_{t,l}}$ and then ${\bf{ x}}_{t,l}$ can be written as
\begin{equation}
\label{x}
\!\!\!\! {\bf{ x}}_{t,l}  =[\hat {\bf{A}}{{\boldsymbol{\hat \omega}}(l)}]^H {\bf{A}}{{\boldsymbol{ \omega}}(l)}  {\bf{s}}_{t,l}\!\!+ \!\![\hat {\bf{A}}{{\boldsymbol{\hat \omega}}(l)}]^H{\bf{ n}} ,
\end{equation}
where ${\bf{ x}}_{t,l} \in \mathbb{C}{^{K \times 1}}$. $\bf{\hat A }$ and ${{\boldsymbol{\hat \omega}}(l)} $ respectively indicate the estimated versions of $\mathbf{A}$ and $\boldsymbol{\omega}(l)$ evaluated using the estimated location parameters. $\bf{\hat A }$ and ${{\boldsymbol{\hat \omega}}(l)} $ are written as  ${\bf{ \hat A}} = \big[{{\bf{ a}}}({\hat \theta _1}) ,...,{{\bf{  a}}}({\hat \theta _K}) ]$ and ${{\boldsymbol{\hat \omega}}(l)}  = \mathrm{diag}(\eta_1(\hat d_1){{\mathrm{e}}^{j2\pi {\hat f_{D,1}}/{f_s}}},...,\eta_K(\hat d_K){{\mathrm{e}}^{j2\pi {\hat f_{D,K}}/{f_s}}} )$, respectively. ${\hat \theta _k}$, $\hat d_k$ and ${\hat f_{D,k}} $ for $k \in \{1,...,K\}$ can be written as ${\hat \theta _K}={ \theta _k}+\Delta { \theta _k}$, $\hat d_k=d_k+\Delta { d_{k}}$ and ${\hat f_{D,k}}={ f_{D,k}}+\Delta { f_{D,k}}$, in which $\Delta { \theta _K}$, $\Delta { d_{k}}$ and $\Delta { f_{D,k}} $ denote the estimation errors of the location parameters.

The $k$th element of ${\bf{ x}}_{t,l}$ in \eqref{x}, that represents the $k$th drone's signal, can be given by
\begin{equation}
	\label{received signal}
	{{ x}}_{t,l,k}\!\! = \!\!\sqrt{{{P}}_{K}}{{\boldsymbol{\hat {h } }^H_k}} {{\boldsymbol{ {h } }_k}} {{s_{t,l,k}}}\!+\!\!\!\!\!\!\sum\limits_{ {i} = 1,{i} \ne k}^K \!\!\!\!\!\!\sqrt{{{P}}_{{i}}}{{\boldsymbol{\hat {h } }^H_k}} {{\boldsymbol{ {h } }_{i}}} {{s_{t,l,{i}}}}\!+ \!{{\boldsymbol{\hat {h } }^H_k}}{\bf{ n}}, 
\end{equation}
where ${\boldsymbol{\hat {h } }_k}$ indicates the estimated channel response of drone $k$, which is composed of the estimated location parameters as $ \vspace{0.05cm}{\boldsymbol{\hat {h } }_k}={ \eta_k}(\hat d_k){{\bf{ a}}}({\hat \theta _k}) {{\mathrm{e}}^{j2\pi (\hat f_{D,k})/{f_s}}}$. ${{\boldsymbol{ {h } }_k}} $ and ${{\boldsymbol{ {h } }_{i}}} $ refer to the actual channel response of drone $k$ and drone ${i}$, which can be represented by using the general expression $\displaystyle {\boldsymbol{ {h } }_p}= \eta_p{{\bf{ a}}}({ \theta _p}) {{\mathrm{e}}^{j2\pi { f_{D,p}}/{f_s}}}$, where $p \in\{1,...,K\}$. Since the estimated location parameters contain estimation errors, ${\boldsymbol{\hat {h } }_k}\in{\mathbb{C}^{N \times 1}}$ can also be written as 
\begin{equation}
\label{h_hat_k}
    {\boldsymbol{\hat {h } }_k}={\eta_k(d_k+\Delta{d_k})}{{\bf{   a}}}({ \theta _k}+\Delta{ \theta _k}) {{\mathrm{e}}^{j2\pi (f_{D,k}+\Delta{f_{D,k}})/{f_s}}}.
\end{equation}

Then the symbol transmitted from the $k$th drone can be detected by using ${{ x}}_{t,l,k}$ in \eqref{received signal}.
It should be noted that due to random noise, $\Delta{d_k}$, $\Delta{ \theta _k} $ and ${\Delta{ f_{D,k}}}$ are random, and so is ${\boldsymbol{\hat {h } }_k}$. It is found that the PDFs of the estimation errors under noise follow a Gaussian distribution, as shown in Fig. \ref{PDF}. In this figure, a BS composed of $N=8$ antennas is employed to estimate the location of a drone, which is located at $({\theta},{f_{D}},d) = [ ({40^ \circ },4000\; \mathrm{Hz},80\; \mathrm{m})]$, in frame $v$ by using the ML-based algorithm with $L=50$ pilots. However, due to noise ${\bf{  n}}$, estimation errors occur. Samples of the estimation errors are collected from a large number of tests at $\mathrm{SNR}= 12\;\mathrm{dB}$, and their PDFs are plotted in Fig. \ref{PDF}. By calculating the mean and variance of the samples and comparing the PDF of the estimation errors to the theoretical Gaussian PDF with the same mean and variance as those of the estimation errors, it is found that the estimation errors follow a Gaussian distribution with zero mean.  

This can be proven according to \cite[Theorem 7.1]{estimation theory}, if the derivatives of the log-likelihood function of the MLE exist and the Fisher information is non-zero, the estimated parameters using MLE follow the Gaussian distribution shown as 
\begin{equation}
    \hat \psi \overset{a}{\sim} \mathcal{N}(\psi, {{F}}^{-1}(\psi)),
\end{equation}
where $\overset{a}{\sim}$ indicates "asymptotically distributed according to" and ${{F}}(\psi)$ represent the fisher information at $\psi$.

The PDF of the estimation errors can be expressed as 
\begin{equation}
\label{PDF Gaussian}
    f(\Delta {\boldsymbol{  \psi}}_k) = \frac{1}{{\sqrt {2\pi } \sigma_\psi }}{{\rm{e}}^{ - \frac{1}{2}{{\left(\frac{\Delta {\boldsymbol{  \psi}}_k}{\sigma_\psi }\right)}^2}}},
\end{equation}
where $\sigma_\psi$ indicates the standard derivation of $\Delta {\boldsymbol{  \psi}}_k=[\Delta { \theta _K}, \Delta { d_{k}}, \Delta { f_{D,k}} ]^T$, which equals to the RMSE as both of them are defined using the same equation as 
\begin{equation}
\label{RMSE_variance}
    \sigma_\psi=\text{RMSE}\overset{\triangle}{=} \sqrt{{\mathbb{E}}[ (\hat \psi_{k}-\psi_k)]^2},
\end{equation}
where $\hat \psi_{k}$ represents the estimated $\psi_k$ . Thus the estimation errors have a variance equal the RMSE of the estimated
parameter.

With this we proceed with the performance analysis in Sec. \ref{performance analysis DTDD} to evaluate the localisation and data decoding performance of the PASCAL system, respectively.

\section{Performance Analysis for PASCAL}
\label{performance analysis DTDD}
\subsection{Cramér-Rao lower
bound (CRLB)}
In this section, the CRLB for the mean squared error (MSE) of the location estimator in Sec. \ref{ML localisation} is presented. To guarantee correspondence with the ML algorithm, the CRLB is derived for the individual localisation within each frame. To obtain the CRLB, the Fisher information matrix (FIM) is calculated first. By considering $K$ drones with each containing three unknown parameters (DOA, range and Doppler frequency), the dimension of the FIM ${\bf{F}}$ is ${\bf{F}} \in \mathbb{C}{^{3K \times 3K}}  $. Hence, the $(i,j)$th submatrix of ${\bf{F}}$ can be given as
\begin{equation}
{\bf{F}}_{i,j} \overset{\triangle}{=}  - \mathbb{E}\Big[\frac{{{\partial ^2}\ln f({{\bf{ y}}_1}|\boldsymbol{\psi} )}}{{\partial \boldsymbol{\psi}_i \partial {\boldsymbol{\psi}_j ^T}}} \Big],
\end{equation}
where  ${{{\partial }\ln f({{\bf{ y}}_1}|\boldsymbol{\psi} )}}/{{\partial \boldsymbol{\psi}_i}} $ and $ {{{\partial }\ln f({{\bf{ y}}_1}|\boldsymbol{\psi} )}}/{{ \partial {\boldsymbol{\psi}_j}}}$ represent the partial derivatives.  ${{\boldsymbol{\psi} }_i}\!=\!{[{{{{\theta}}}_i },d_i,{{{f_{D,i}}} }]^T}$  and ${{\boldsymbol{\psi} }_j}\!=\!{[{{{{\theta}}}_j },d_j,{{{f_{D,j}}} }]^T}$ for $i,j \in \{1,...,K\}$. Since ${{\bf{ y}}_1}$ follows the multivariate Gaussian distribution, the Slepian-Bangs formula \cite{Slepian--Bangs formula} can be invoked to simplify ${\bf{F}}_{i,j}$ as 
\begin{equation}
\label{F_ij}
{{\bf{F}}_{i,j}} = \mathrm{tr}\left[{{\bf{\Gamma }}^{ - 1}}\frac{{\partial {\bf{\Gamma }}}}{{\partial {{\boldsymbol{\psi} }_i}}}{{\bf{\Gamma }}^{ - 1}}\frac{{\partial {\bf{\Gamma }}}}{{\partial {{\boldsymbol{\psi} }_j}}}\right] + 2{{\Re}} \left[\frac{{\partial {\boldsymbol{\mu} ^H}}}{{\partial {{\boldsymbol{\psi} }_i}}}{{\bf{\Gamma }}^{ - 1}}\frac{{\partial \boldsymbol{\mu} }}{{\partial {{\boldsymbol{\psi} }_j}}}\right],
\end{equation}
where $\mathrm{tr}( \cdot )$ indicates the trace function and ${{\Re}}( \cdot ) $ denotes the real part of the input argument.

By noting that $\displaystyle {{\partial {{\bf{\Gamma }}}}}/{{\partial {{\boldsymbol{\psi }}_i}}}=$ $\displaystyle {{\partial {{\bf{\Gamma }}}}}/{{\partial {{\boldsymbol{\psi }}_j}}}=0$, ${\bf{F}}_{i, j}$ in \eqref{F_ij} can be further simplified to
\begin{equation}
{{\bf{F}}_{i,j}} =  2{{\Re}} \left[\frac{{\partial {\boldsymbol{\mu} ^H}}}{{\partial {{\boldsymbol{\psi} }_i}}}{{\bf{\Gamma }}^{ - 1}}\frac{{\partial \boldsymbol{\mu} }}{{\partial {{\boldsymbol{\psi} }_j}}}\right],
\end{equation}
where $ \displaystyle {{\partial {\boldsymbol{\mu}}}}/{{\partial {{\boldsymbol{\psi} }_i}}}$ and $\displaystyle {{\partial {\boldsymbol{\mu}}}}/{{\partial {{\boldsymbol{\psi} }_j}}} $ can be represented by using a general expression as $\displaystyle {{\partial {\boldsymbol{\mu}}}}/{{\partial {{\boldsymbol{\psi} }_k}}}$, in which  $\displaystyle {{ {\partial \boldsymbol{\mu}}}}/{{\partial {{\boldsymbol{\psi} }_k}}} = [{{ \partial {\boldsymbol{\mu}}}}/{\partial{{\theta}}_k },{\partial{ {\boldsymbol{\mu}}}}/{{\partial {{{d}}_k }}},{\partial{ {\boldsymbol{\mu}}}}/{{\partial {{{f_{D}}}_k }}}]^T  $, where
\begin{subequations}
\label{small}
\begin{align}
\label{muphi0}
    \displaystyle \frac{{\partial \boldsymbol{\mu} }}{{\partial \theta}_k} &\!\!=\!\!\sqrt{{{P}}_{k}}{\Big\{ \Big[ \frac{{\partial {{\bf{ a}}}({\theta _k})}}{{\partial {{{\theta}_k} }}}{{{\omega}}_k}(1)\Big]^T,..., \Big[\frac{{\partial {{\bf{ a}}}({\theta _k})}}{{\partial {{{\theta}_k} }}}{{{\omega}}_k}(l)\Big]^T\Big\} }^T,
\end{align}
\vspace{-0.4cm}
\begin{align}
\label{muf0}
   \displaystyle \frac{{\partial \boldsymbol{\mu} }}{{\partial {{{d}}_k }}} &\!\!= \!\!
 \sqrt{{{P}}_{k}}{ \Big\{\Big[{{\bf{ a}}}({\theta _k}) \frac{{ \partial{{{\omega}}_k}(1)}}{{\partial {{{d}}_k }}}\Big]^T,..., \Big[{{\bf{ a}}}({\theta _k})\frac{{\partial {{{\omega}}_k}(l)}}{{\partial {{{d}}_k }}}\Big]^T\Big\}}^T,
\end{align}
\vspace{-0.4cm}
\begin{align}
\label{muf0}
   \displaystyle \frac{{\partial \boldsymbol{\mu} }}{{\partial {{{{f_{D}}}_k }_k}}} &\!\!= \!\!
 \sqrt{{{P}}_{k}}{ \Big\{\Big[{{\bf{ a}}}({\theta _k}) \frac{{ \partial{{{\omega}}_k}(1)}}{{\partial {{{{f_{D}}}_k }}}}\Big]^T,..., \Big[{{\bf{ a}}}({\theta _k})\frac{{\partial {{{\omega}}_k}(l)}}{{\partial {{{f_{D}}}_k }}}\Big]^T\Big\}}^T,
\end{align}
\end{subequations}
and ${{\partial {{\bf{ a}}}({\theta _k})}}/{{\partial {{{\theta}_k} }}}$ can be obtained as ${{\partial {{\bf{ a}}}({\theta _k})}}/{{\partial {{{\theta}_k} }}}={\boldsymbol{\Lambda}}_\theta{{\bf{ a}}}({\theta _k})$, in which ${\boldsymbol{\Lambda}}_\theta= \mathrm{diag}\{\Lambda_1,...,\Lambda_N\} $ and $\Lambda_n=-j2\pi(n-1)d_0\cos \theta_k/\lambda $ for $n\in \{1,...,N\}$. In addition, ${{\partial {{{\omega}}_k}(l)}}/{{\partial {d_k}}}=-1/d_k{{{\omega}}_k}(l)$ and ${{\partial {{{\omega}}_k}(l)}}/{{\partial {f_{D,k}}}}=j 2\pi l/f_s{{{\omega}}_k}(l)$ can also be evaluated by using simple mathematical calculations.  

\subsection{Conditional SER for MPSK}
\label{conditional SER MPSK}
In conventional channel estimation, the estimated channel is usually assumed to follow a complex Gaussian distribution \cite{channelestimation_Guerreiro,channelestimation_Raeesi,channelestimation_Wang}. However, this assumption may not be applicable to our case. By observing the ${\boldsymbol{\hat {h } }_k}$ in \eqref{h_hat_k}, it can be found that ${\boldsymbol{\hat {h } }_k}$ is composed of multiple random parameters (i.e., $\Delta{d_k}$, $\Delta{ \theta _k} $ and ${\Delta{ f_{D,k}}}$), each of which follows a Gaussian distribution. In addition, the random parameter $\Delta{ \theta _k} $ is contained in the elements corresponding to different antennas of ${\boldsymbol{\hat {h } }_k}$. Therefore, it is difficult to determine the distribution of the channel in our case. Furthermore, the derivation of the average SER is not straightforward as well not only because the distribution of the channel is unknown but also due to the fact that ${\boldsymbol{\hat {h } }_k}$ is contained in both the desired signals' part and the interference part of the preprocessed signal in \eqref{received signal}, which causes a significant correlation. 

 In order to obtain the average SER for $M$-ary Phase Shift Keying (MPSK), the conditional SER given the estimation errors of location parameters $\Delta {\boldsymbol{  \psi}} = {[{\Delta {{ \theta }}},{\Delta{{ d}}},{\Delta{{ f_{D}}}}]^T}$, which can be calculated by using $\Delta {\boldsymbol{  \psi}}=\hat {\boldsymbol{  \psi}}-{\boldsymbol{  \psi}} $, should be derived first.  To begin, by considering all possible phase-shifted signal combinations transmitted by different drones, the conditional SER can be calculated by using
\vspace{-0.1cm}
\begin{equation}
\label{conditional SER}
    {P_{{e}}|\Delta {\boldsymbol{  \psi}}}  \!\!=\!\!\sum\limits_{{m_1},...,{m_K} \in {S_1}}\!\! \frac{  P_{{e}}|\{\Delta {\boldsymbol{  \psi}},\beta\} }{M^{K}} ,
    \vspace{-0.1cm}
\end{equation}
where $S_1\!=\!\{1,2,...,M\}$ and $M$ indicates the number of possible phases in modulating the signals using MPSK.  $\beta=\{{s_{t,l,1}} = {{\bf{s}}_{t,l,1}}(m_1),...,{s_{t,l,K}} = {{\bf{s}}_{t,l,K}}(m_K) \} $ represents the phase-shifted signal combination, in which, for instance, drone $1$ and drone $K$ are transmitting ${{\bf{s}}_{t,l,1}}(m_1)$ and ${{\bf{s}}_{t,l,K}}(m_K) $, respectively for ${m_1},...,{m_K} \in {S_1}$. ${{\bf{s}}_{t,l,1}}$,..., ${{\bf{s}}_{t,l,K}}$ can be expressed by using a general expression as   ${{\bf{s}}_{t,l,k}}$, which denotes the $t$th signal vector with all possible phases transmitted by drone $k$ in the $l$th subframe and ${{\bf{s}}_{t,l,k}}$ is written as  ${{\bf{s}}_{t,l,k}} = [1,...,{{\rm{e}}^{{{j2\pi (M - 1)}}/{M}}}]$.

\begin{figure}[ptb]
	\centering
	\includegraphics [scale=0.46]{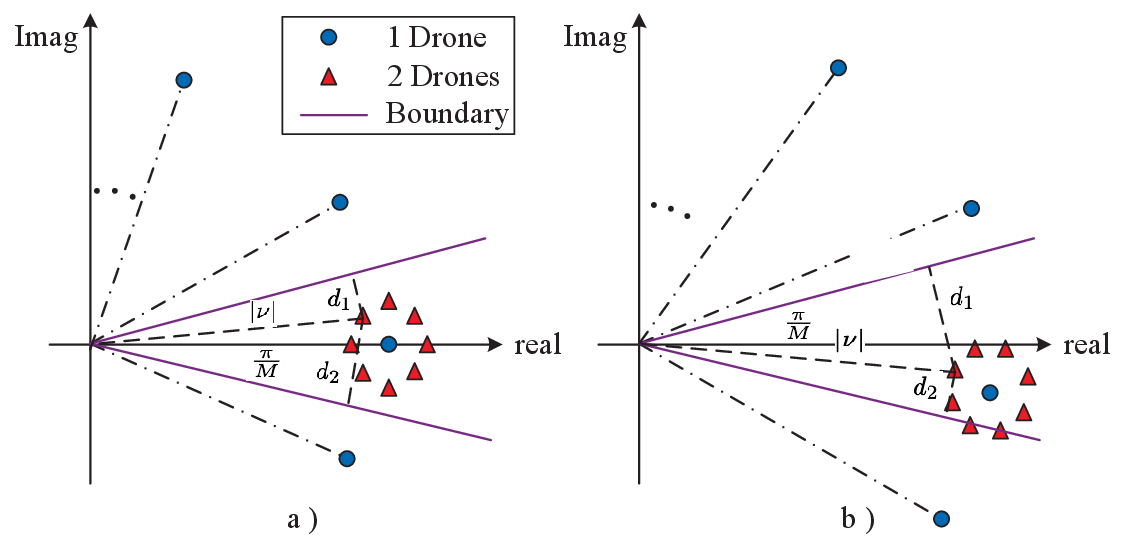}
	\caption{Illustrations of union bound method for MPSK a) with perfect localisation b) with imperfect localisation for two drones.}%
	\label{union bound}

\end{figure}
 
By noting that the only random parameter in ${{ x}}_{t,l,k}$ given $\Delta {\boldsymbol{  \psi}}$ is the noise vector ${\bf{ n}}$, the PDF of ${{ x}}_{t,l,k}$ conditioned on $\Delta {\boldsymbol{  \psi}}$ can be given by 
\begin{equation}
\label{conditional PDF}
\displaystyle f({{ x}}_{t,l,k}|\Delta {\boldsymbol{  \psi}} ) = \frac{1}{{ {{\Gamma }}}_x}{{\mathrm{e}}^{ - \frac{{{({{ x}}_{t,l,k} - {\mu}_x )}^*}({{ x}}_{t,l,k} - {\mu}_x )}{{{\Gamma }}_x}}},
\end{equation}
where ${\mu}_x $ and ${{ {{\Gamma }}}_x} $ refers to the mean and variance of ${{ x}}_{t,l,k}$.  ${\mu}_x $ can be written as
\begin{equation}
 {\mu}_x=\sqrt{{{P}}_{K}}{{\boldsymbol{\hat {h } }^H_k}} {{\boldsymbol{ {h } }_k}} {{s_{t,l,k}}}\!+\!\!\!\!\!\!\sum\limits_{ i = 1,i \ne k}^K \!\!\!\!\!\!\sqrt{{{P}}_{i}}{{\boldsymbol{\hat {h } }^H_k}} {{\boldsymbol{ {h } }_{i}}} {{s_{t,l,{i}}}}    
\end{equation}

In addition, ${{ {{\Gamma }}}_x} $ can be calculated by using ${{ {{\Gamma }}}_x}={\mathbb{E}}[({{ x}}_{t,l,k}-{\mu}_x)^*({{ x}}_{t,l,k}-{\mu}_x)]$ and the result can be simplified to
\begin{equation}
    {{ {{\Gamma }}}_x}=\sum\limits_{ i = 1}^N\sum\limits_{ j = 1}^N{{{\hat {h } }_{k,i}}}{{{\hat {h } }^*_{k,j}}}{\mathbb{E}}[{{ n}}^*_i{{ n}}_j],
\end{equation}
where ${{{\hat {h } }_{k,i}}}$ and ${{{\hat {h } }_{k,j}}} $ indicate the $i$th and $j$th element of ${{\boldsymbol{\hat {h } }_{k}}}$, while ${{ n}}_i$ and ${{ n}}_j $ refer to the $i$th and $j$th element of ${\bf{ n}}$. By performing some simple algebraic manipulations and noting that ${{{\hat {h } }_{k,i}}}{{{\hat {h } }^*_{k,j}}}={ \eta^2_k}(\Delta{d_k})$ for $i=j$, ${{ {{\Gamma }}}_x}$ can be further simplified to ${{ {{\Gamma }}}_x} =N{ \eta^2_k}(\Delta{d_k}) \sigma^2 $ and $\eta_k$ is a function of $\Delta{d_k} $.

\begin{figure*}[!b]
\hrulefill
 \setcounter{equation}{37} 
 \vspace{-0.2cm}
 \begin{align}
\label{cossin}
&\prod_{g_1=1}^{G_1}\prod_{g_2=1}^{G_2} \cos{ x_{g_1}}\sin{ x_{g_2}}=
    \begin{cases}   &\displaystyle \!\!\sum\limits_{{e_1,...,e_{G_1+G_2}}\in S_1}\frac{(-1)^{\frac{G_2}{2}}\cos(\sum\limits_{g = 1}^{G_1+G_2}  {e_g}{x_{g}})\prod\limits_{g = 1}^{G_2}e_g}{2^{G_1+G_2}},
    \ \ \  \text{if \textit{$G_2$} is even }
    \vspace{-0.3cm}
    \\
&\displaystyle \!\!\sum\limits_{{e_1,...,e_{G_1+G_2}}\in S_1}\frac{(-1)^{\frac{G_2-1}{2}}\sin(\sum\limits_{g = 1}^{G_1}   {e_g}{x_{g}}+{e_{G_1+1}}[{x_{G_1+1}}+\!\!\sum\limits_{g = G_1+2}^{G_1+G_2}   {e_g}{x_{g}}])\prod\limits_{g = 1}^{G_2}e_g}{2^{G_1+G_2}},
   \ \ \   \text{if \textit{$G_2$} is odd }
   \vspace{-0.3cm}
    \end{cases}
\end{align}  
 \vspace{-0.3cm}
\end{figure*}

Based on the conditional PDF in \eqref{conditional PDF}, $P_{{e}}|\{\Delta {\boldsymbol{  \psi}},\beta\}$ can be derived using the union bound method, which is illustrated in Fig. \ref{union bound} for the example of two drones. It is worth mentioning that even if Fig. \ref{union bound} uses an example of two drones, our derivation for the conditional SER is general for multiple drones.  In Fig. \ref{union bound}, the blue circles and red triangles denote the constellation points considering drone 1 and the constellation points considering drone 1 and drone 2, in which the signal transmitted from drone 2 is considered as the interference signal when we are decoding the symbols from drone 1. $\tilde d_1$ and $\tilde d_2$ refer to the distance from the constellation points to the boundary lines without the influence of the noise term. $|\nu|$ indicates the distance between the constellation point and the origin of the coordinate axis.  

In Fig. \ref{union bound}, the union bound methods with the perfect localisation case and the imperfect localisation case for deriving the conditional MPSK SER are compared. When the perfect localisation case is considered, the only interference affecting the data decoding process is the noise $\bf{n}$. However, both $\bf{n}$ and the location errors can affect the data decoding process when the imperfect localisation case is considered. Interestingly, it can be found that the constellation points are rotated in the imperfect localisation case compared to those of the perfect localisation case due to the localisation errors. In addition, the distance from the constellation points to the origin of the coordinate axis is also affected. In order to derive the conditional MPSK SER with imperfect localisation, the probability that the constellation points lie outside the boundary lines under the influence of the noise $\bf{n}$ is required, which can be obtained by calculating the probability that the distance from the constellation points to the two boundary lines is respectively greater than $\tilde d_1$ and $\tilde d_2$. Since ${{ x}}_{t,l,k}$ given $\Delta {\boldsymbol{  \psi}}$ follow the Gaussian distribution, as shown in \eqref{conditional PDF}, $P_{{e}}|\{\Delta {\boldsymbol{  \psi}},\beta\}$ can be evaluated as 
 \setcounter{equation}{20} 
\begin{equation}
 \label{conditional SER1}
    P_{{e}}|\{\Delta {\boldsymbol{  \psi}},\beta\}=\frac{1}{{\sqrt {\pi \Gamma_x} }}\int_{ \tilde d_1  }^\infty {{\mathrm{e}^{\frac{{ - {{{z}  }^2}}}{\Gamma_x}}}} d{z}+\frac{1}{{\sqrt {\pi \Gamma_x} }}\int_{\tilde d_2  }^\infty {{\mathrm{e}^{\frac{{ - {{{z}  }^2}}}{{\Gamma_x}}}}} d{z},
\end{equation}
where $\tilde d_1$ and $\tilde d_2$ can be respectively given by 
\begin{subequations}
\label{dd}
    \begin{align}
    \tilde d_1 \!\!&=\!\! \sin \left(\frac{\pi }{M} - \arg (\nu )\right)|\nu | \nonumber \\
   \!\! &= \!\! \left\{\sin \frac{\pi }{M} \cos(\arg(\nu))-\cos \frac{\pi }{M} \sin(\arg(\nu))\right\}|\nu |, 
    \end{align}
    \begin{align}
     \tilde d_2 \!\!&=\!\! \sin \left(\frac{\pi }{M}+\arg (\nu ) \right)|\nu | \nonumber \\ 
     \!\! &= \!\! \left\{\sin \frac{\pi }{M} \cos(\arg(\nu))+\cos \frac{\pi }{M} \sin(\arg(\nu))\right\}|\nu |,
    \end{align}
\end{subequations}
where $\nu=\sqrt{{{P}}_{k}}{{\boldsymbol{\hat {h } }^H_k}} {{\boldsymbol{ {h } }_k}} {{{\bf{s}}_{t,l,k}}}(m_k)\!+\!\!\!\!\sum\limits_{ i = 1,i \ne k}^K \!\!\!\!\sqrt{{{P}}_{i}}{{\boldsymbol{\hat {h } }^H_k}} {{\boldsymbol{ {h } }_{i}}} {{{\bf{s}}_{t,l,i}}}(m_{i})$, in which $m_k, m_{i} \in \{m_1,...,m_K\}$. To facilitate the subsequent analysis, we denote $\nu$ as $\nu=\!\!\sum\limits_{ p = 1}^K \!\!\sqrt{{{P}}_{p}}{{\boldsymbol{\hat {h } }^H_k}} {{\boldsymbol{ {h } }_p}} {{{\bf{s}}_{t,l,p}}}(m_p)$, in which $p\in\{1,...,K\}$.

By replacing the integrals in \eqref{conditional SER1} with Q functions, substituting ${{ {{\Gamma }}}_x} =N{ \eta^2_k}(\Delta{d_k}) \sigma^2 $ in \eqref{conditional SER1} and performing some algebraic manipulations, $P_{{e}}|\{\Delta {\boldsymbol{  \psi}},\beta\}$ in \eqref{conditional SER1} can also be given by
\setcounter{equation}{22} 
\begin{equation}
\vspace{-0.1cm}
\label{Pe Qfunction}
    P_{{e}}|\{\Delta {\boldsymbol{  \psi}},\beta\}=Q\!\left(\! \frac {\sqrt{2 } \tilde d_1}{\sqrt{{N{ \eta^2_k}(\Delta{d_k}) \sigma^2}}} \!\right)\!+\!Q\!\left(\! \frac {\sqrt{2 } \tilde d_2}{\sqrt{{N{ \eta^2_k}(\Delta{d_k}) \sigma^2}}}\! \right)\!,
\end{equation}

To simplify the notation in the subsequent analysis, we define $ Q_1= Q\left( \frac {\sqrt{2 } \tilde d_1}{\sqrt{{N{ \eta^2_k}(\Delta{d_k}) \sigma^2}}} \right)$ and $ Q_2= Q\left( \frac {\sqrt{2 } \tilde d_2}{\sqrt{{N{ \eta^2_k}(\Delta{d_k}) \sigma^2}}} \right) $. By substituting the result of $P_{{e}}|\{\Delta {\boldsymbol{  \psi}},\beta\}$ into \eqref{conditional SER}, the conditional SER for MPSK can be obtained. 

\subsection{Average SER for MPSK}
\label{average SER}
Based on the conditional SER derived in Sec. \ref{conditional SER MPSK}, the average SER for MPSK can be calculated by using
\setcounter{equation}{23} 
\vspace{-0.1cm}
\begin{align}
\label{P_e}
    {P_{{e}}}  \!\!=\!\!\sum\limits_{{m_1},...,{m_K} \in {S_1}}\!\! \frac{  {\mathbb{E}}[P_{{e}}|\{\Delta {\boldsymbol{  \psi}},\beta\} ]}{M^{K}}.
    \vspace{-0.2cm}
\end{align}

By substituting the derivation result of $P_{{e}}|\{\Delta {\boldsymbol{  \psi}},\beta\}$ in \eqref{Pe Qfunction} into \eqref{P_e}, ${P_{{e}}}  \!\!=\!\!\sum\limits_{{m_1},...,{m_K} \in {S_1}}\!\! \frac{  {\mathbb{E}}[Q_1 ]+{\mathbb{E}}[Q_2 ]} {M^{K}} \vspace{-0.1cm}$ can be obtained. Then, by substituting the value of $\tilde d_1$ and $\tilde d_2$ in \eqref{dd} into $Q_1$ and $Q_2$, an expression for ${\mathbb{E}}[Q_{\ell} ]$ for $\ell\in\{1,2\}$ can be obtained, which is  
\begin{equation}
\label{E_Ql1}
  \!  {\mathbb{E}}[Q_{\ell} ]\!\!=\!\!{\mathbb{E}}\!\!\left[\!Q\!\! \left(\!\! \frac {\!\sqrt{2 } (\!\sin\!\! \frac{\pi }{M} \!\cos(\arg(\!\nu\!))\!\!+\!\!(\!-\!1\!)^{\ell} \!\cos\!\! \frac{\pi }{M}\! \sin(\arg(\!\nu\!))|\nu |)}{\sqrt{{N{ \eta^2_k}(\Delta{d_k}) \sigma^2}}}\! \!\right)\! \right]\!,
\end{equation}
where $\arg(\nu)$ can be given by
\begin{equation}
\arg(\nu) = 
\begin{cases}
  \arctan\left(\frac{\nu_y}{\nu_x}\right), & \text{if } \nu_x > 0 \\
  \arctan\left(\frac{\nu_y}{\nu_x}\right) + \pi, & \text{if } \nu_x < 0 \text{ and } \nu_y \geq 0 \\
  \arctan\left(\frac{\nu_y}{\nu_x}\right) - \pi, & \text{if } \nu_x < 0 \text{ and } \nu_y < 0 \\
  \frac{\pi}{2}, & \text{if } \nu_x = 0 \text{ and } \nu_y > 0 \\
  -\frac{\pi}{2}, & \text{if } \nu_x = 0 \text{ and } \nu_y < 0 \\
  \text{undefined}, & \text{if } \nu_x = 0 \text{ and } \nu_y = 0 \\
\end{cases}    
\end{equation}
where $\nu_x$ and $\nu_y$ represent the real and image part of $\nu$.

Using $\cos (\arctan (x)) = \frac{1}{{\sqrt {1 + {x^2}} }}$ and $\sin (\arctan (x)) = \frac{x}{{\sqrt {1 + {x^2}} }} $, where $x$ indicates the input of $\arctan (\cdot) $, and performing some algebraic operations, a simplified expression for ${\mathbb{E}}[Q_{\ell} ]$  in \eqref{E_Ql1} can be obtained, which can be written as 
    \begin{equation}
    \label{EQ_ell}
{\mathbb{E}}[Q_{\ell}]  \!\!= \!\!
\begin{cases}
  {\mathbb{E}}\left[Q\left( \frac {\sqrt{2 } \tilde \nu}{{\sqrt{{N{ \eta_k(\Delta{d_k})} \sigma^2}}}} \right)\right],  
  & \text{if } \nu_x  \neq  0 \vspace{0.1cm} \\
  {\mathbb{E}}\left[Q \left( {\frac{(-1)^{{\ell}} \sqrt{2} {\cos \frac{\pi }{M}}\nu_y}{{\sqrt{{N { \eta_k}(\Delta{d_k}) \sigma^2}}}}} \right)\right], & \text{if } \nu_x = 0\! \text{ and }\! \nu_y\neq 0   \vspace{0.1cm} \\
  \text{undefined}, & \text{if } \nu_x = 0\! \text{ and }\! \nu_y = 0 \\
\end{cases} 
\end{equation}
where $ \tilde \nu=\sin \frac{\pi }{M} \nu _x + (-1)^{\ell}\cos \frac{\pi }{M}\nu _y$. 

\begin{figure*}[!b]
\hrulefill
\setcounter{equation}{39} 
\begin{align} 
\label{E 8}
 {\mathbb{E}}{_8} =& 
 \begin{cases}
 \displaystyle \ \ \sum\limits_{{e_1},...,{e_{q_1}} \in {S_1}} {\frac{(-1)^{\frac{q_1}{2}}}{2^{q_1}}\underbrace{{\mathbb{E}}[\cos  ({C_2}{C_3} {\Delta {\theta _k}}  +  2\pi{C_4}+2\pi{C_{5}}\Delta f_{D,k})]}_{{{\mathbb{E}}}_{10}}\prod\limits_{g = 1}^{q_1}e_g},
   \ \ \ \ \ \ \ \ \ \text{if \textit{$q_1$} is even }   \\
 \displaystyle \ \ \sum\limits_{{e_1},...,{e_{q_1}} \in {S_1}} \! \!\!\!\!{\frac{(-1)^{\frac{q_1-1}{2}}}{2^{q_1}}\underbrace{{\mathbb{E}}[\sin  ({C_2}{C_3} {\Delta {\theta _k}}  +  2\pi{C_4}+2\pi{C_{5}}\Delta f_{D,k})]}_{{{\mathbb{E}}}_{11}}\prod\limits_{g = 1}^{q_1}e_g},
   \ \ \ \ \ \ \ \ \   \text{if \textit{$q_1$} is odd }
 \end{cases}
\end{align}
\end{figure*}

In order to simplify the notation,  ${\mathbb{E}}\left[Q\left( \frac {\sqrt{2 } \tilde \nu}{{\sqrt{{N { \eta^2_k(\Delta{d_k})} \sigma^2}}}} \right)\right]$ and ${\mathbb{E}}\left[Q \left( {\frac{(-1)^{{\ell}} \sqrt{2} {\cos \frac{\pi }{M}}\nu_y}{{\sqrt{{N { \eta^2_k(\Delta{d_k})}\sigma^2}}}}} \right)\right]$ can be denoted by using ${\mathbb{E}}_1$ and ${\mathbb{E}}_2$, respectively. To simplify ${\mathbb{E}}_1$ and ${\mathbb{E}}_2$ further, $\eta_k(\Delta{d_k})$ can be extracted from $\tilde \nu$ and $\nu_y$  and thus $ \tilde \nu_1=\frac{\nu}{\eta_k(\Delta{d_k})}$ and $\nu_{y1}=\frac{\nu_y}{\eta_k(\Delta{d_k})} $ can be obtained. $\tilde \nu_1$ can also be written as
 \setcounter{equation}{27} 
\begin{equation}
\label{tildenu1}
\tilde \nu_1=\sin \frac{\pi }{M} \nu _{x1} + (-1)^{\ell}\cos \frac{\pi }{M} \nu _{y1},
\end{equation}
where $\nu _{x1}$ can be calculated by using $\nu _{x1}=\frac{\nu _{x}}{ \eta_k(\Delta{d_k})}$. 

By substituting the value of ${{\boldsymbol{\hat {h } }^H_k}}$,  ${{\boldsymbol{ {h } }_p}}$, ${{{\bf{s}}_{t,l,p}}}(m_p)$ into $\nu=\!\!\sum\limits_{ p = 1}^K \!\!\sqrt{{{P}}_{p}}{{\boldsymbol{\hat {h } }^H_k}} {{\boldsymbol{ {h } }_p}} {{{\bf{s}}_{t,l,p}}}(m_p)$ we can find $\nu _{x} $ and $\nu _{y}  $, where $\nu _{x} $ and $\nu _{y}  $ denote the real and image parts of $\nu $. Using $\nu _{x1}=\frac{\nu _{x}}{ \eta_k(\Delta{d_k})}$ and $\nu _{y1}=\frac{\nu _{y}}{ \eta_k(\Delta{d_k})} $, $\nu _{x1} $ and $\nu _{y1}  $ can be obtained as
\setcounter{equation}{28}
\begin{subequations}
\label{nu_xy}
\begin{align}
   {\nu}_{x1} =\mathop \sum \limits_{p = 1}^K  \sum\limits_{i = 1}^{N } \eta_p\sqrt{P_p}{\cos{(  j2\pi((i- 1)d_0{\Theta _p}+\Phi)) } }, 
\end{align}
\vspace{-0.4cm}
\begin{align}
   {\nu}_{y1} =\mathop \sum \limits_{p = 1}^K  \sum\limits_{i = 1}^{N } \eta_p\sqrt{P_p}{\sin{(  j2\pi((i- 1)d_0{\Theta _p}+\Phi)) } }, 
\end{align}    
\end{subequations}
where $\Phi={ \frac{{{m_p}\! -\! 1}}{M} \!+\! \frac{{{f_{D,p}}\! -\! {f_{D,k}}\! -\! \Delta {f_{D,k}}}}{{{f_s}}}} \vspace{0.1cm}$ and $ {\Theta _p} = [\sin ({\theta _k} + \Delta {\theta _k}) - \sin {\theta _p}]/\lambda$. By using the following small-angle approximation 
\begin{align}
    \sin ({ \theta _k}+\Delta{ \theta _{k}})&=\sin { \theta _k}\cos{\Delta{ \theta _{k}}}+\sin{\Delta{ \theta _{k}}}\cos { \theta _k}\nonumber \\
    &\simeq \sin { \theta _k}+\Delta{ \theta _{k}}\cos { \theta _k},
\end{align}
$ {\Theta _p}$ can be approximated as $ \simeq [\sin {\theta _k} - \sin {\theta _p}+\Delta{ \theta _{k}}\cos { \theta _k}]/\lambda$. 
\vspace{0.2cm}

Then by substituting $ \tilde \nu_1=\frac{\nu}{\eta_k(\Delta{d_k})}$ and $\nu_{y1}=\frac{\nu_y}{\eta_k(\Delta{d_k})} $ into ${\mathbb{E}}_1$ and ${\mathbb{E}}_2$, ${\mathbb{E}}_1$ and ${\mathbb{E}}_2$ can be simplified as ${\mathbb{E}}\left[Q\left( \frac {\sqrt{2 } \tilde \nu_1}{{\sqrt{{N \sigma^2}}}} \right)\right]$ and ${\mathbb{E}}\left[Q \left( {\frac{(-1)^{{\ell}} \sqrt{2} {\cos \frac{\pi }{M}}\nu_{y1}}{{\sqrt{{N\sigma^2}}}}} \right)\right]$. It can be found that ${ \eta_k(\Delta{d_k})}$ is cancelled out in ${\mathbb{E}}_1$ and ${\mathbb{E}}_2$. Since ${\nu}_{x1}$ and ${\nu}_{y1}$ in \eqref{nu_xy} don't contain ${ \eta_k(\Delta{d_k})}$, ${\tilde \nu}_{1}$ and ${\nu}_{y1}$ are not a function of ${ \eta_k(\Delta{d_k})}$.  As a consequence, we can conclude that the estimation errors for range have no effect on the average SER for MPSK using MRC.

Based on the conclusion that the estimation errors of range do not affect the average SER, ${\mathbb{E}}_1$ and ${\mathbb{E}}_2$ can be evaluated as 
\setcounter{equation}{30} 
\begin{subequations}
\label{E1 E2}
\begin{align}
   {\mathbb{E}}_1 \!\!&=\!\!\!\! \int_{\Delta {f_{D,\min }}}^{\Delta {f_{D,\max }}}\!\!\int_{ - \pi }^\pi\!  \! \!\!Q\!\left(\! \frac {\sqrt{2 } \tilde \nu_1}{\sqrt{{N\sigma^2}}} \!\right)\!  f(\Delta\! {\theta _k})f(\Delta \!{f_{D,k}})d\Delta {\theta _k} d\Delta {f_{D,k}}, \\
    {\mathbb{E}}_2 \!\!&=\!\! \!\!\int_{\Delta {f_{D,\min }}}^{\Delta {f_{D,\max }}}\!\! \!\! \int_{ - \pi }^\pi \!\! \!\!{Q \!\left( \!{\!\frac{ \sqrt{2} C_1\nu_{y1}}{\sqrt{N\sigma^2}}} \!\right)\!} f(\Delta\! {\theta _k})f(\Delta \!{f_{D,k}})d\Delta{\theta _k} d\Delta{f_{D,k}},
\end{align}    
\end{subequations}
where $C_1=(-1)^{{\ell}}{\cos\! \frac{\pi }{M}}$, $f(\Delta {\theta_k} )$ and $f(\Delta {f_{D,k}})$ denote the PDF of $\Delta {\theta _k}$ and $\Delta {f_{D,k}} $, respectively, which can be obtained by using \eqref{PDF Gaussian}.

\subsection{An Taylor Approximation approach for Average SER}
\label{Approximation}
Due to the fact that the Q functions in ${\mathbb{E}}_1$ and ${\mathbb{E}}_2$ are composed of the sum of trigonometric functions that contain random parameters following Gaussian distributions ($\tilde \nu_1$ and $\nu_{y,1}$ contain $\Delta {\theta _k}$ and $\Delta {f_{D,k}}$, as shown in \eqref{nu_xy}), there is no closed-form solution for ${\mathbb{E}}_1$ and ${\mathbb{E}}_2$. As a consequence, the $R$th Taylor polynomial of the Q function is invoked to approximate $Q(x)$ in ${\mathbb{E}}_1$ and ${\mathbb{E}}_2$, where $x$ denotes the input. To improve the approximation performance and increase the convergence speed, the approximation of $Q(x)$ for $x$ at the point $x_0$ is employed, in which $x_0$ indicates the perfect case of $x$ when there are no estimation errors for DOA and Doppler frequency (i.e., $\Delta {\theta _k}=0$, $\Delta {f _{D,k}}=0$). It is worth mentioning that this perfect estimation for location parameters can be obtained when $\mathrm{SNR}$ is very large. In addition, since $\Delta {\theta _k}$ and $\Delta {f _{D,k}}$ distribute around zero with a variance that decreases considerably with the increase of SNR, this kind of Taylor approximation is very accurate even if $R$ is small. Then the Taylor approximation of $Q(x)$ at $x_0$ can be denoted by
\vspace{-0.1cm}
\begin{equation}
\label{Taylor approximation}
    Q(x)\simeq\sum\limits_{{r} = 0}^{{R}} {\frac{Q^{(r)}(x_0){{{(x- {x_{0}})}^{{r}}}}}{{{r}!}}},
    \vspace{-0.1cm}
\end{equation}
where $Q^{(r)}(x_0)$ indicates the $r$th derivative of the Q function evaluated at $x_0$ and $Q^{(0)}(x_0)=Q(x_0)$. While for $r>0$, $Q^{(r)}(x_0)$ can be expressed as
\begin{equation}
    Q^{(r)}(x_0)=-\frac{1}{(\sqrt{2})^{r}\sqrt{\pi}}{\rm{e}}^{-\frac{x^2_0}{2}}(-1)^{r+1}H_{r-1}\big(\frac{x_0}{\sqrt{2}}\big),
\end{equation}
where $\displaystyle H_{r-1}\big({x_0}/{\sqrt{2}}\big)$ denotes the $(r-1)$th Hermite polynomial at $\displaystyle {x_0}/{\sqrt{2}}$. 
\begin{figure*}[!b]
\hrulefill
\setcounter{equation}{41} 
\begin{align} 
\label{E 9}
 {\mathbb{E}}{_9} = & 
 \begin{cases}
 \displaystyle \sum\limits_{{e_1},...,{e_{{q_1}}} \in {S_1}} {\frac{(-1)^{\frac{q_1-q_2}{2}}}{{2^{{q_1} }}} \underbrace{{\mathbb{E}}[\cos  ({C_2}{C_3} {\Delta {\theta _k}}  +  2\pi{C_4}+2\pi{C_{5}}\Delta f_{D,k})]}_{{\mathbb{E}}_{10}}\prod\limits_{g = q_2+1}^{q_1}e_g},
   \ \ \  \text{if \textit{$q_1-q_2$} is even }  \vspace{-0.2cm} \\
  \displaystyle \sum\limits_{{e_1},...,{e_{q_1}} \in {S_1}}\!\!\!\!\! {\frac{(-1)^{\frac{q_1-q_2-1}{2}}}{{2^{q_1}}}\underbrace{{\mathbb{E}}[\sin  ({C_2}{C_{6} {\Delta {\theta _k}} + 2\pi{C_{7}}+2\pi{C_{8}}\Delta f_{D,k})]}}_{{\mathbb{E}}_{12}}\prod\limits_{g = q_2+1}^{q_1}e_g},
   \ \ \  \text{if \textit{$q_1-q_2$} is odd }
 \end{cases}
\end{align}
\vspace{-0.2cm}
\setcounter{equation}{43} 
\begin{align}
\label{C 7}
    &{C_{7}} =  \sum\limits_{{g} = 1}^{{q_2}}  e_g \Phi_g + e_{q_2+1}\bigg\{\sum\limits_{{g} = q_2+2}^{{q_1}}  e_g \Phi_g + \frac{({i_{{{q_2+1}}}} - 1)d_0(\sin {\theta _k} - \sin {\theta _{{p_{{q_2+1}}}}})}{\lambda} + \frac{{{m_{{p_{{q_2+1}}}}} - 1}}{M} + \frac{{{f_{D,{p_{{q_2+1}}}}} - {f_{D,k}}}}{{{f_s}}}\bigg\},
\end{align}
\end{figure*}
${\mathbb{E}}_1$ and ${\mathbb{E}}_2$ can be approximated as
 \setcounter{equation}{33} 
 \begin{equation}
 \label{E 1 E 2}
    {\mathbb{E}}_{\ell}\simeq \sum\limits_{{r} = 0}^{{R}}\! \frac{(\sqrt{2})^r Q^{(r)}(x_{{\ell},0})(C^r_1)^{\ell-1}{\mathbb{E}}_{\ell+2}}{({\sqrt{N\sigma^2}})^r r!},
\end{equation}  
where $\displaystyle x_{\ell,0}={\sqrt{2 } \tilde \nu_{1,0}}/{\sqrt{{N\sigma^2}}}$ if ${\ell}=1$ and $ \displaystyle x_{\ell,0}= {\sqrt{2 } C_1 \nu_{y1,0}}/{\sqrt{{N\sigma^2}}}$ if ${\ell}=2$, in which $\vspace{0.1cm} \tilde \nu_{1,0}$ and $\nu_{y1,0}$ indicate the prefect case for $\tilde \nu_1$ and $\nu_{y1}$ when there is no localisation errors. In addition, $\vspace{0.1cm}{\mathbb{E}}_3={\mathbb{E}}[(\tilde \nu_1- {\tilde \nu_{1,0}})^r]$ and ${\mathbb{E}}_4={\mathbb{E}}[( \nu_{y1}- { \nu_{y1,0}})^r] $. 

To evaluate ${\mathbb{E}}_3$ and ${\mathbb{E}}_4$, the binomial theorem needs to be used to expand the polynomials in ${\mathbb{E}}_3$ and ${\mathbb{E}}_4$. Afterwards, ${\mathbb{E}}_3$ and ${\mathbb{E}}_4$ can be written as
\vspace{-0.2cm}
\begin{equation}
\label{E 3 E 4}
 {{\mathbb{E}}_{\ell+2}} = \sum\limits_{{q_1}=0}^{r} {\left(\! {\begin{array}{*{20}{c}}
{{r}}\\
{{q_1}}
\end{array}} \!\right)}  C_{\ell+2}^{{{r-q_1}}}{\mathbb{E}}_{\ell+4} ,  
\vspace{-0.2cm}
\end{equation}
where ${\mathbb{E}}_{\ell+4}$ can be given by ${\mathbb{E}}_5= \displaystyle {\mathbb{E}}[ \tilde \nu^{{q_1}}_1]   $ and ${\mathbb{E}}_6=\displaystyle {\mathbb{E}}[  \nu^{{q_1}}_{y1}]$, $C_3=-\tilde \nu_{1,0} \vspace{0.1cm}$ for $\ell=1$ and $C_4=-\nu_{y1,0} $ for $\ell=2$. By substituting $\displaystyle \tilde \nu_1=\sin \frac{\pi }{M} \nu _{x1} + (-1)^{\ell}\cos \frac{\pi }{M}\nu _{y1} \vspace{0.05cm}$ into ${\mathbb{E}}_5$ and applying the binomial theorem to ${\mathbb{E}}_5$, ${\mathbb{E}}_5$ can also be given as 
\begin{align}
\label{E 5}
  \!\!\!\! {\mathbb{E}}_5&= \displaystyle {\mathbb{E}} \left[\left(\sin \frac{\pi }{M} \nu _{x1} + (-1)^{\ell}\cos \frac{\pi }{M}\nu _{y1}\right)^{{q_1}}\right]\nonumber \vspace{0.15cm}  \\
  &=\sum\limits_{{q_2}=0}^{q_1} {\left(\!\! {\begin{array}{*{20}{c}}
{{q_1}} \\
{{q_2}}
\end{array}} \!\!\right)} \left(\sin \frac{\pi}{M}\right)^{q_2}  \left((-1)^{\ell}\cos \frac{\pi }{M}\right)^{q_1-q_2} {\mathbb{E}}_{7},
\end{align}
where ${\mathbb{E}}_{7}=\displaystyle {\mathbb{E}}  [\nu^{{q_2}}_{x1} \nu^{{q_1-q_2}}_{y1}]   $.

To calculate ${\mathbb{E}}_6\displaystyle ={\mathbb{E}}[  \nu^{{q_1}}_{y1}]$ and ${\mathbb{E}}_7=\displaystyle {\mathbb{E}}  [\nu^{{q_2}}_{x1} \nu^{{q_1-q_2}}_{y1}]$, the product-to-sum identities of the trigonometric formulas are employed here to transform the product of trigonometric functions into a single cosine function or a single sine function to simplify the subsequent analysis. The trigonometric formulas are 
\setcounter{equation}{36}
\begin{subequations}
 \begin{align}
\label{cos}
    &\prod_{g=1}^{G} \cos{x_g}=\frac{1}{2^{G}}\sum\limits_{{e_1,...,e_{G}}\in S_1}{\cos(\sum\limits_{g = 1}^{G}  {e_g}{x_{g}})},
    \\
    &\prod_{g=1}^{G} \!\sin{\! x_g}\!\!=\!\!
    \begin{cases}   &\!\!\!\!\!\!\!\sum\limits_{{e_1,...,e_{G}}\in S_1}\!\!\!\frac{(\!-1\!)^{\frac{G}{2}}\!\cos(\sum\limits_{g = 1}^{G}  {e_g}{x_{g}})\!\!\prod\limits_{g = 1}^{G}e_g}{2^{G}},
    \ \ \  \text{if \textit{G} is even }
    \\
    \label{sin}
&\!\!\!\!\!\!\!\sum\limits_{{e_1,...,e_{G}}\in S_1}\!\!\!\frac{(\!-1\!)^{\frac{G\!-\!1}{2}}\!\sin(\sum\limits_{g = 1}^{G}   {e_g}{x_{g}})\!\!\prod\limits_{g = 1}^{G}e_g}{2^{G}},
   \ \ \   \text{if \textit{G} is odd }
    \end{cases}
\end{align}   
\end{subequations}
and \eqref{cossin} at the bottom of page \pageref{cossin}, in which $G$, $G_1$ and $G_2$ indicate the upper limits of the products, $x_g$, $x_{g_1}$ and $x_{g_2}$ denote the inputs, and ${S_1}=\{-1,1\}^{G}$. Using the above trigonometric formulas, substituting the values of ${\nu}_{x1}$ and ${\nu}_{y1}$ shown in \eqref{nu_xy} into the calculations, and performing some algebraic simplification, ${\mathbb{E}}_6$ and ${\mathbb{E}}_7$ can also be expressed by a general expression as
\setcounter{equation}{38} 
\begin{equation}
\label{E 6 E 7}
    {\mathbb{E}}{_{\ell+5}} \!\!=  \!\!\!\!\!\!\!\! \sum\limits_{{p_{1}}, \cdots ,{p_{q_1}} \in {S_2}}  \sum\limits_{{i_1}, \cdots ,{i_{q_1}} \in {S_3}}   {\prod\limits_{g = 1}^{q_1} \eta_{p_g} \sqrt{P_{p_{g}}}}{\mathbb{E}}_{{\ell+7}},
\end{equation}
where $S_2 \in \{1,...,K\}$, $S_3 \in \{1,...,N\}$, and $\ell \in \{1,2\}$. ${\mathbb{E}}{_8}$ is shown in \eqref{E 8} at the bottom of page \pageref{E 8}, where $C_2=2\pi d_0 \cos \theta_k/\lambda$, $C_3=\sum\limits_{g = 1}^{q_1}  {e_g}({i_{g}}-1)$, $C_4=\sum\limits_{g = 1}^{q_1}  {e_g}\Phi_g $ and $C_5=-\frac{1}{f_s}\sum\limits_{g = 1}^{q_1}  {e_g}$. In addition, $\Phi_g$ can be written as
\setcounter{equation}{40} 
\begin{equation}
\label{C 4}
    \Phi_g\!\!=\!\! {(i_g \!-\!
    1)(\sin {\theta _k} \!-\! \sin {\theta _{p_g}})\frac{d_0}{\lambda} \!+\!\frac{{{m_{p_g}} \!-\! 1}}{M} \!+ \!\frac{{{f_{D,{p_g}}} \!-\! {f_{D,k}} }}{{{f_s}}}},
\end{equation}
where it is worth mentioning that the notation $ i_g$ is employed to differentiate $i$ in different summation operations and the same applies for ${p_{g}}$, $\eta_{p_{g}}$, $P_{p_{g}}$, $\sin \theta_{p_g}$, $m_{p_{g}}$, $f_{D,p_{g}}$ and $e_g$. 

Similar to ${\mathbb{E}}_8$, ${\mathbb{E}}_9$ is shown in \eqref{E 9} at the bottom of page \pageref{E 9}, where 
 \setcounter{equation}{42} 
 \begin{align}
{C_6 }\!\! &=\!\! \sum\limits_{g = 1}^{q_2}  {e_g}({i_{g}}\!-\!1) \! +\!{e_{q_2+1}}[({i_{q_2+1}}\!-\!1)\!\!+ \!\!\!\!\!\!\sum\limits_{g ={q_2+2} }^{q_1}{e_g}({i_{g}}\!-\!1)] , 
\end{align}   
where the two summation notations indicate the first $q_2$ summations and the last $(q_1-q_2-1)$ summations, respectively. Similarly, $C_7$ is shown in \eqref{C 7} at the bottom of page \pageref{C 7}, where the definition of $\Phi_g$ is shown in \eqref{C 4}. ${C_8 }$ can be represented as  
\setcounter{equation}{44} 
\begin{equation}
{C_{8} } = -\frac{1}{f_s}\bigg[\sum\limits_{g = 1}^{q_2}  {e_g}  + {e_{q_2+1}}\Big(1+\!\!\sum\limits_{g ={q_2+2} }^{q_1}{ e_g}\Big)\bigg].
\end{equation}

To calculate ${\mathbb{E}}{_{8}}$ and ${\mathbb{E}}{_{9}}$ in \eqref{E 8} and \eqref{E 9}, we need to evaluate the double integrals in ${\mathbb{E}}{_{10}}$, ${\mathbb{E}}{_{11}}$ and ${\mathbb{E}}{_{12}}$ (i.e., ${\mathbb{E}}{_{10}}$, ${\mathbb{E}}{_{11}}$ and ${\mathbb{E}}{_{12}}$ contain two random variables), the inner integrals with respect to $\Delta {\theta _k}$ should be calculated first by taking $\Delta {f_{D,k}} $ as a constant. Afterwards, the results of the inner integrals will be integrated with respect to $\Delta {f_{D,k}} $. By substituting the expression for $f(\Delta {\theta_k})$, which can be obtained  using \eqref{PDF Gaussian}, into ${\mathbb{E}}{_{10}}$ and ${\mathbb{E}}{_{11}}$, the inner integrals in ${\mathbb{E}}{_{10}}$ and ${\mathbb{E}}{_{11}}$ can be obtained and written as
\begin{subequations}
    \begin{align}
     {I}_{10}&=\displaystyle
  \!\!{\frac{1}{\sqrt {2\pi } \sigma_\theta }} \!\!\int_{-\pi}^{\pi }  {\! \cos  ({C_2}{C_3}\Delta {\theta _k}\!+\!C_{9}) } {{\rm{e}}^{\! - \frac{1}{2}{{\left(\frac{{\Delta {\theta _k}}}{\sigma_\theta }\right)}^2}}}\!\!d\Delta \theta_k,  
    \end{align}
    \begin{align}
    {I}_{11}\!&=\displaystyle
  \!\!{\frac{1}{\sqrt {2\pi } \sigma_\theta }} \!\! \int_{-\pi}^{\pi }  {\! \sin  ({C_2}{C_{3}}\Delta {\theta _k}\!+\!C_{9}) } {{\rm{e}}^{ - \frac{1}{2}{{\left(\frac{{\Delta {\theta _k}}}{\sigma_\theta }\right)}^2}}}\!\!d\Delta \theta_k,   
    \end{align}
   \begin{align}
    \!{I}_{12}\!&=\displaystyle
  \!\!{\frac{1}{\sqrt {2\pi } \sigma_\theta }}\!\! \int_{-\pi}^{\pi }  {\! \sin  ({C_2}{C_{6}}\Delta {\theta _k}\!+\!C_{10}) } {{\rm{e}}^{ - \frac{1}{2}{{\left(\frac{{\Delta {\theta _k}}}{\sigma_\theta }\right)}^2}}}\!\!d\Delta \theta_k, \! 
  \end{align}
\end{subequations}
where $C_{9}= 2\pi{C_4}+2\pi{C_{5}}\Delta f_{D,k}$ and $C_{10}=  2\pi{C_7}+2\pi{C_{8}}\Delta f_{D,k}$. 

Afterwards, by using Euler's formula and performing some mathematical calculations, ${I}_{10}$ and ${I}_{11}$ can be obtained and their results can be denoted by using a general expression as
\begin{equation}
\label{I10 I11}
    {I}_{\ell+9}=\frac{1}{ 2\sqrt {2\pi } \sigma_\theta(j)^{\ell-1} }({\rm{e}}^{jC_{9}}{I_{1}}+(-1)^{\ell+1}{\rm{e}}^{-jC_{9}}{I_{2}}),
\end{equation}
and ${I}_{12}$ evaluated as
\begin{align}
\label{ I 12}
{I}_{12}=
      \frac{1}{j 2\sqrt {2\pi } \sigma_\theta }({\rm{e}}^{jC_{10}}{I_{3}}-{\rm{e}}^{-jC_{10}}{I_{4}}),
\end{align}
where $I_1$, $I_2$, $I_3$ and $I_4$ can be expressed by using the general expression
 \begin{equation}
    {I_5} = \int_{ - \pi}^{\pi} {{\rm{e}}^{ \pm j{C_{11}{\Delta {\theta _k}}{\rm{ - }}\frac{{\rm{1}}}{{{\rm{2}}\sigma _\theta ^2}}{\Delta {\theta ^2_k}}}}} d {\Delta {\theta _k}},
\end{equation}
where $C_{11} \in \{C_2 C_3,C_2 C_6  \}$. The complete derivation of $I_5$ can be found in Appendix A. By using the derived closed-form expression of ${I_5}$, the results of $I_1$, $I_2$, $I_3$ and $I_4$ can be obtained, which are then substituted into \eqref{I10 I11} and \eqref{ I 12} to obtain the closed-form expressions of ${I_{10}}$, ${I_{11}}$ and ${I_{12}}$. 

Thereafter, the result of ${I_{10}}$, ${I_{11}}$ and ${I_{12}}$ can be integrated with respect to $\Delta {f_{D,k}} $ to calculate ${\mathbb{E}}_{10}$, ${\mathbb{E}}_{11}$ and ${\mathbb{E}}_{12} $ as
\begin{subequations}
    \begin{align}
     {\mathbb{E}}_{10}
    &=\int_{\Delta {f_{D,\min }}}^{\Delta {f_{D,\max }}}{I}_{10}{{\rm{e}}^{ - \frac{1}{2}{{\left(\!\!\frac{{\Delta {f _{D,k}}}}{\sigma_{f_D} }\!\!\right)}^2}}}d\Delta {f_{D,k}},  
    \end{align}
    \vspace{-0.4cm}
    \begin{align}
     {\mathbb{E}}_{11}&=
   \int_{\Delta {f_{D,\min }}}^{\Delta {f_{D,\max }}}{I}_{11}{{\rm{e}}^{ - \frac{1}{2}{{\left(\!\!\frac{{\Delta {f _{D,k}}}}{\sigma_{f_D} }\!\!\right)}^2}}}d\Delta {f_{D,k}}.   
    \end{align}
    \vspace{-0.4cm}
    \begin{align}
    {\mathbb{E}}_{12}&=
   \int_{\Delta {f_{D,\min }}}^{\Delta {f_{D,\max }}}{I}_{12}{{\rm{e}}^{ - \frac{1}{2}{{\left(\!\!\frac{{\Delta {f _{D,k}}}}{\sigma_{f_D} }\!\!\right)}^2}}}d\Delta {f_{D,k}}.
    \end{align}
\end{subequations}

By using Euler's formula again and performing some mathematical manipulation, ${\mathbb{E}}_{10}$ and ${\mathbb{E}}_{11}$ can be shown to be
\begin{equation}
\label{E10 E11}
{\mathbb{E}}_{\ell+9}\!\!=\!\!\frac{1}{{(j)^{\ell+1}4\pi } \sigma_\theta \sigma_{f_D} }({\rm{e}}^{j2 \pi C_{7}}{I_1}{I_6}+(-1)^{\ell+1}{\rm{e}}^{-j2 \pi C_{7}}{I_2}{I_7}),
\end{equation}
where $\ell \in \{1,2\}$ and $C_7$ can be found in \eqref{C 7} at the bottom of page \pageref{C 7}. Similarly, ${\mathbb{E}}{_{12}}$ can be evaluated as
\begin{align}
  \label{E12}
   {\mathbb{E}}_{12}=\frac{1}{{j4\pi } \sigma_\theta \sigma_{f_D} }({\rm{e}}^{j2 \pi C_{7}}{I_3}{I_8}-{\rm{e}}^{-j2 \pi C_{7}}{I_4}{I_9}),
\end{align}
where $I_1$, $I_2$, $I_3$ and $I_4$ have been derived and thus they are constants in \eqref{E10 E11} and \eqref{E12}. $I_6$, $I_7$, $I_8$ and $I_9$ can be expressed by using the general expression
 \begin{equation}
    {I_{13}} = \int_{\Delta {f_{D,\min }}}^{\Delta {f_{D,\max }}} {{\rm{e}}^{ \pm j{C_{12}{\Delta {f_{D,k }}}{\rm{ - }}\frac{{\rm{1}}}{{{\rm{2}}\sigma _{f_D} ^2}}{\Delta {f^2_{D,k }}}}}} d {{\Delta {f_{D,k }}}},
\end{equation}
where $C_{12} \in \{2 \pi C_{5},2 \pi C_{8}  \}$. The complete derivation of $I_{13}$ can be found in Appendix A. By using the derived closed-form expression of $I_{13}$, $I_6$, $I_7$, $I_8$ and $I_9$ can be obtained, which are then substituted into \eqref{E10 E11} and \eqref{E12} to obtain the closed-form expressions of ${{\mathbb{E}}_{10}}$, ${{\mathbb{E}}_{11}}$ and ${{\mathbb{E}}_{12}}$. 

\begin{table}[t]
\small
\label{R_1_9b}
\caption{Simulation Parameters}
\vspace{-0.2cm}
\centering{}
\newsavebox{\tablebox}
\begin{lrbox}{\tablebox}
\begin{tabular}{|c|c|c|c|c|c|}
\hline
\textbf{Param.}                    &   \textbf{Value} &\textbf{Param.}                    &   \textbf{Value} &\textbf{Param.}                    &   \textbf{Value}
\\
\hline
${\theta}_1$     & $20^ \circ$  & ${d}_1$     & $80 \; \mathrm{m}$  & ${f}_{D,1}$     & $ 2\mathrm{kHz}$ 
\\
\hline
${\theta}_2$     & $40^ \circ$  & ${d}_2$     & $80 \; \mathrm{m}$  & ${f}_{D,2}$     & $ 4\mathrm{kHz}$ 
\\
\hline
$v_1$     & $3.4 \; \mathrm{m/s}$  & $v_2$     & $8.4 \; \mathrm{m/s}$  &  $K$    & 2
\\
\hline
$T$      &100    &   $f_s$       & 100 $\mathrm{kHz}$ &$\lambda$     & 1.6  $\mathrm{mm}$\\
\hline
\end{tabular}
\end{lrbox}
\scalebox{0.9}{\usebox{\tablebox}}
\label{sim par}
\vspace{-0.3cm}
\end{table}

It is worth mentioning that the derived ${{\mathbb{E}}_{10}}$, ${{\mathbb{E}}_{11}}$ and ${{\mathbb{E}}_{12}}$ are a function of the RMSE of the estimated location parameters since the closed-form expressions of $I_5$ and $I_{13}$ in \eqref{I5 I13} contains $\sigma_\psi$, which denotes the RMSE, as shown in \eqref{RMSE_variance}. Consequently, the derived SER is a function of the RMSE of the estimated location parameters in our PASCAL system.

Consequently, the derived results are employed to obtain the average SER. First, the derived ${{\mathbb{E}}_{10}}$, ${{\mathbb{E}}_{11}}$ and ${{\mathbb{E}}_{12}}$ are substituted into \eqref{E 8} and \eqref{E 9} to obtain ${{\mathbb{E}}_{8}}$ and ${{\mathbb{E}}_{9}}$, which are substituted into \eqref{E 6 E 7} to obtain ${{\mathbb{E}}_{6}}$ and ${{\mathbb{E}}_{7}}$ with $\ell \!\in\! \{1,2\}$. By using the result of ${{\mathbb{E}}_{7}}$, ${{\mathbb{E}}_{5}}$ can be calculated according to \eqref{E 5}. Thereafter, ${{\mathbb{E}}_{5}}$ and ${{\mathbb{E}}_{6}}$ can be substituted into \eqref{E 3 E 4} to calculate ${{\mathbb{E}}_{3}}$ and ${{\mathbb{E}}_{4}}$. Finally, ${{\mathbb{E}}_{1}}$ and ${{\mathbb{E}}_{2}}$ can be obtained by substituting  ${{\mathbb{E}}_{3}}$ and ${{\mathbb{E}}_{4}}$ into \eqref{E 1 E 2}, where ${{\mathbb{E}}_{1}}={\mathbb{E}}\left[Q\left( \frac {\sqrt{2 } \tilde \nu}{{\sqrt{{N{ \eta_k(\Delta{d_k})} \sigma^2}}}} \right)\right]$ and ${{\mathbb{E}}_{2}}={\mathbb{E}}\left[Q \left( {\frac{(-1)^{{\ell}} \sqrt{2} {\cos \frac{\pi }{M}}\nu_y}{{\sqrt{{N { \eta_k}(\Delta{d_k}) \sigma^2}}}}} \right)\right]$. According to \eqref{EQ_ell}, \vspace{0.1cm} ${\mathbb{E}}[Q_{\ell}]$ can be calculated by using ${\mathbb{E}}_1$ and ${\mathbb{E}}_2$, which \vspace{0.1cm} can be substituted into ${P_{{e}}}  \!\!=\!\!\sum\limits_{{m_1},...,{m_K} \in {S_1}}\!\! \frac{  {\mathbb{E}}[Q_1 ]+{\mathbb{E}}[Q_2 ]}{M^{K}}$ to obtain ${P_{{e}}} $.
\vspace{-0.2cm}

\section{Numerical Results}
\label{results}
This section presents the simulation results and analytical results to evaluate the performance of the PASCAL system, where RMSE and SER are the localisation performance metric and communication performance metric, respectively. Two drones are considered in our simulations, and the simulation parameters can be found in Table \ref{sim par}, where Param. indicates parameter. In addition, without loss of generality, the transmit power of the signals from drone 1 and drone 2 is considered to be the same, and thus $P_1=P_2$. The ML-based localisation algorithm in Sec. \ref{ML localisation} is employed to estimate the locations of drones in each frame, while MRC is employed to preprocess the received signal to enhance the signal quality before data decoding.  In each simulation point, 1000 Monte Carlo tests are conducted.

\begin{figure}[ptb]
\centering
{\includegraphics  [height=2.0in, width=3.2in]{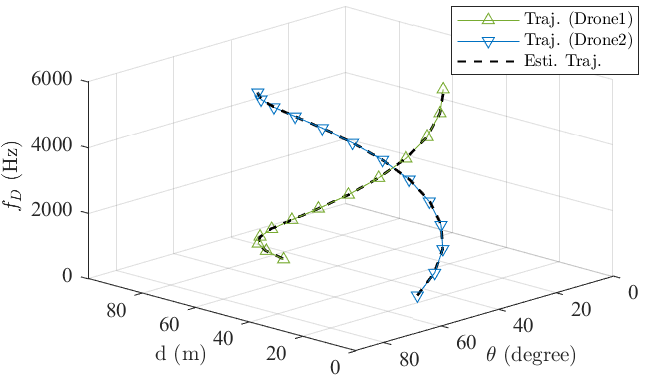}}
\caption{The trajectory estimations of two drones.}
\color{black}
\label{Trajectory}%
\end{figure}

In Fig. \ref{Trajectory}, the trajectories of drone 1 and drone 2 are estimated by using the ML-based localisation algorithm in Sec. \ref{ML localisation}. In the simulation, a BS composed of $N=8$ antennas is employed and $L=50$ pilots are employed by the ML-based algorithm to achieve localisation of the two drones at $\mathrm{SNR}= 12\;\mathrm{dB}$. It is worth mentioning that the trajectories of drone 1 and drone 2 by continually estimating the locations of the drones when the drones are moving. In Fig. \ref{Trajectory}, Traj. and Esti. Traj. denote real trajectory and estimated trajectory, respectively. It can be found that the estimated trajectory and the real trajectory match perfectly for both drone 1 and drone 2. For instance, the estimated locations of drone 1 and drone 2 by using MLE are $({\theta},{f_{D}},d) = [({44.704^ \circ },\! {0.102\; \mathrm{Hz} },\!75.481\; \mathrm{m}), ({45.103^ \circ },\!0.203\; \mathrm{Hz},\!24.955\; \mathrm{m})]$ when the drones are located at $({\theta},{f_{D}},d) = [({45.000^ \circ },\!{0.000\; \mathrm{Hz} },\!75.000\; \mathrm{m}), ({45.000^ \circ },\!0.000\; \mathrm{Hz},\!25.000\; \mathrm{m})]$. The RMSE is $(0.222^ \circ,0.161 \mathrm{Hz},0.342 \mathrm{m} )$ in this case by calculating the average RMSEs of the estimated location parameters corresponding to two drones, which indicates the efficiency of the ML-based algorithm in localisation. 

 \begin{figure*}[t]
\centering 
{\includegraphics  [height=1.43in, width=7in]{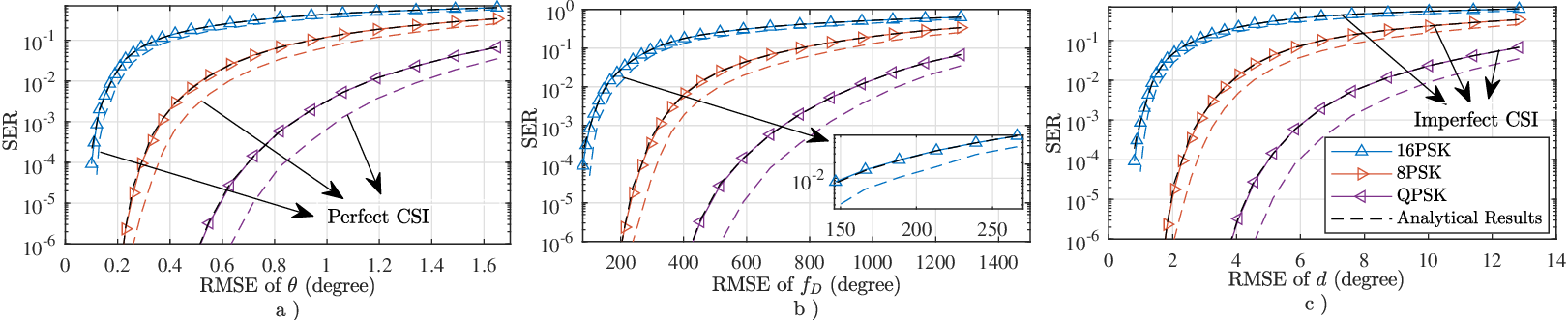}}
\caption{ The SERs with RMSEs of a) DOA $\theta$, b) Doppler frequency $f_D$, c) Range $d$} 
\label{tradeoff}
\end{figure*}

 \begin{figure*}[t]
\centering 
{\includegraphics  [height=2.27in, width=7in]{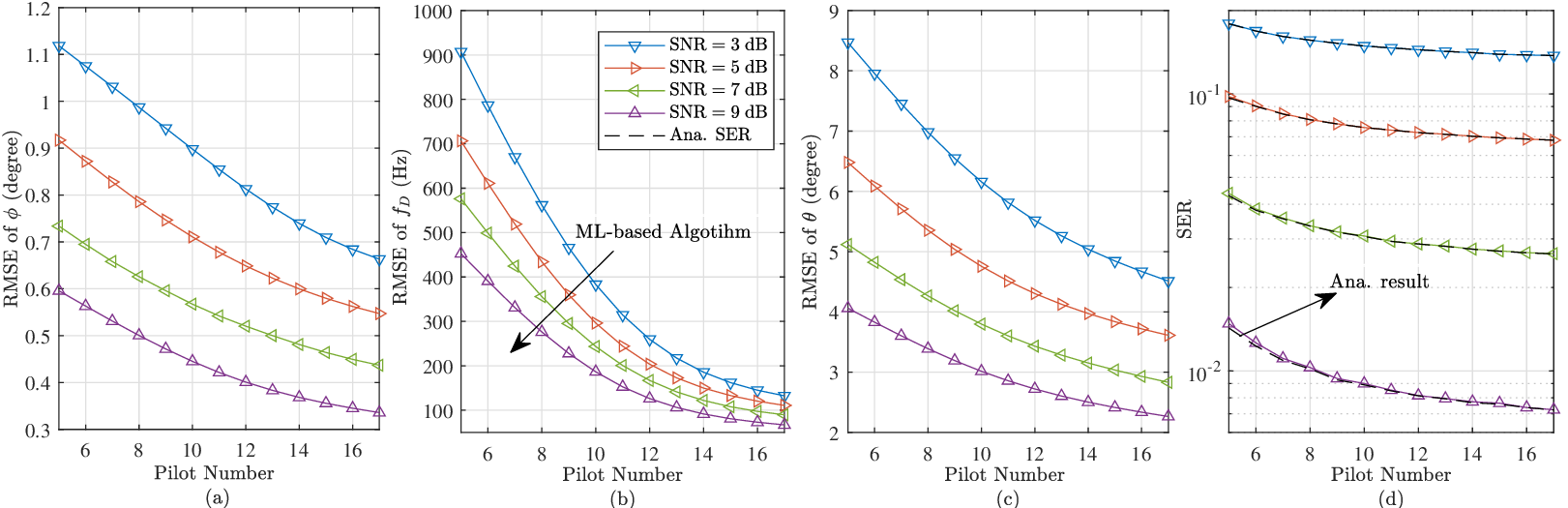}}
\caption{ The a) RMSE of DOA $\theta$, b) RMSE of Doppler frequency $f_D$, c) RMSE of Range $d$, d) SER with the increase of the number of pilots} 
\label{pilot number}
\end{figure*}

  \begin{figure}[ptb]
\centering
{\includegraphics  [height=2.20in, width=2.9in]{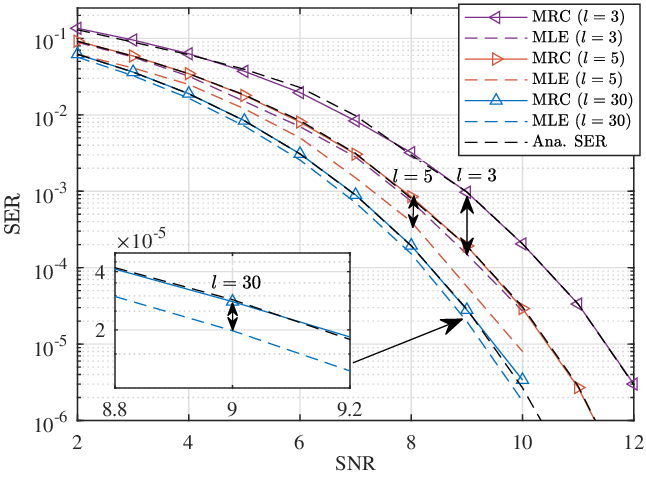}}
\caption{The performance gap between MRC and MLD.}
\color{black}
\label{MLD}
\vspace{-0.4cm}
\end{figure}

In Fig. \ref{tradeoff}, the data SER is plotted against the RMSE in location estimation. The figure clearly shows the synergy relationship between the localisation part and the communication part of the PASCAL system. In the simulation, a BS composed of $N=6$ antennas is employed to localise and serve drone 1 and drone 2, and $L=5$ pilots are contained in each frame. In addition, three modulation types including QPSK, 8PSK and 16PSK are considered. To obtain the curves in Fig. \ref{tradeoff}, SNR has been changed from $0 \;\mathrm{dB}$ to $24\; \mathrm{dB}$. As can be observed from the results, the localisation accuracy has a positive effect on SER since a smaller RMSE corresponds to a lower SER, in which the RMSE and SER represent the average values of the RMSEs and SERs for two drones. This conclusion applies to the curves generated by using the three modulation schemes. This phenomenon can be attributed to the fact that the channel information is obtained by using the estimated location parameters in each frame. However, there are some estimation errors of location parameters due to the presence of AWGN, which leads to erroneous channel estimation. As a consequence, an improved localisation accuracy leads to a more accurate channel estimation and thus reduces SER. The performance gap between the imperfect CSI with the perfect CSI, which is obtained by combining the perfect knowledge of CSI with the received signal by using the MRC technique and serves as a benchmark, is also shown in Fig. \ref{tradeoff}. For instance, the SER gap between the perfect CSI case and the imperfect CSI case is $1.3 \times 10^{-4}$. In addition, analytical results and simulated results of SER match perfectly for all modulation types, where the $R=6$th Taylor approximation in \eqref{Taylor approximation} is employed to obtain the analytical result.   

In Fig. \ref{pilot number}, both the communication and localisation performance of the PASACL system with the increase of pilot number is demonstrated. Without loss of generality, we employ the localisation for drone 1 and drone 2 in frame $v$ as an example to indicate the localisation performance of the PASACL system. Since the data decoding of different subframes in frame $v$ are based on a varying number of pilot signals, it is essential to investigate the impact of the number of pilots on them. In the simulation, we consider a BS consisting of $N=6$ antennas. In addition, 8PSK is used as the modulation method. As can be observed from Fig. \ref{pilot number}, both the localisation and communication accuracy improve with the increase in the number of pilot signals. This indicates that even if increasing the pilot number may increase communication latency, it has a positive effect on both the localisation accuracy and communication reliability of the proposed PASACL system. This phenomenon is due to the fact that the localisation accuracy is enhanced by using more pilots, which results in a more accurate acquisition of channel information and a better data decoding result. In addition, the analytical results obtained by using $R=6$th approximation orders in \eqref{Taylor approximation} and the simulation results match very well. 

 \begin{figure*}[t]
\centering 

{\includegraphics  [height=2.27in, width=7in]{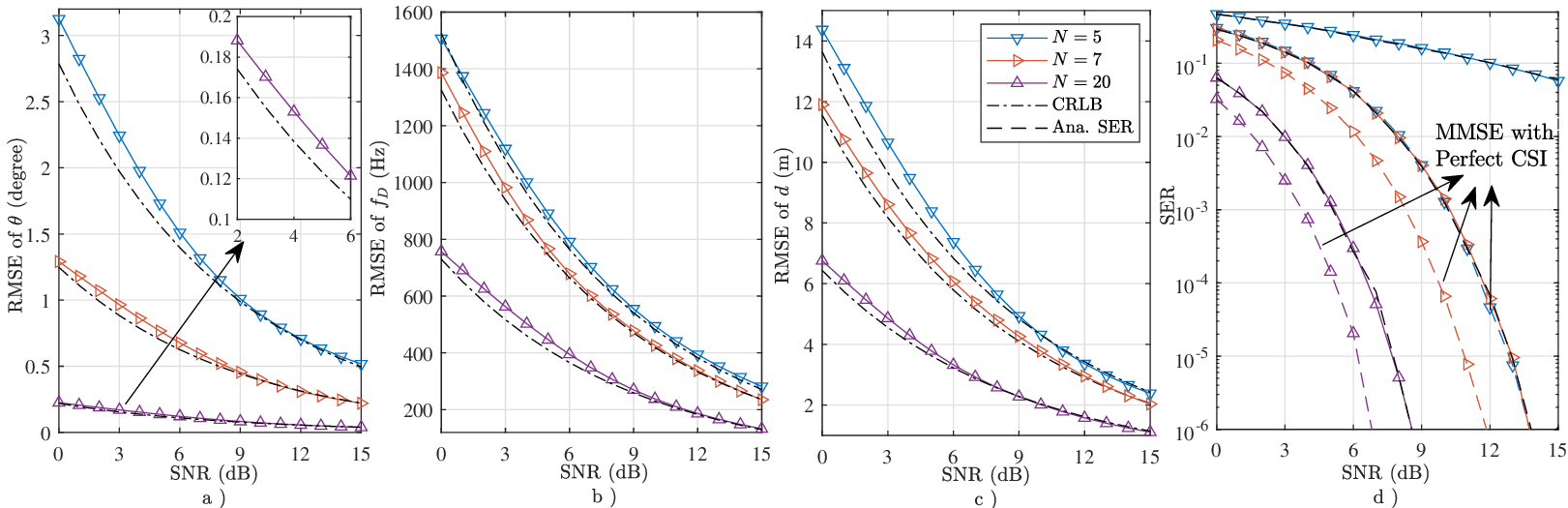}}

\caption{ The a) RMSE of DOA $\theta$, b) RMSE of Doppler frequency $f_D$, c) RMSE of Range $d$, d) SER with the increase of SNR} 
\label{benchmark}
\vspace{-0.2cm}
\end{figure*}

 \begin{figure}[ptb]
\centering 
{\includegraphics  [height=1.29in, width=2.79in]{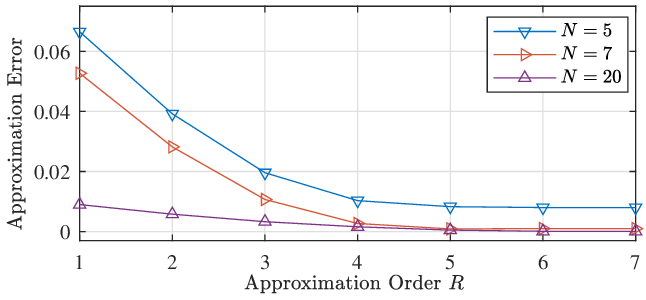}}

\caption{The approximation performance with the increase of approximation order $L$.} 
\vspace{-0.2cm}
\label{approximation error}
\end{figure}

In Fig. \ref{MLD}, the performance of MRC is compared to that of the optimal detector MLE \cite{MLD} in different subframes in frame $v$. Notably, MLE is not referred to as the ML-based localisation algorithm in Sec. \ref{ML localisation}, instead, it is conducted by estimating location parameters including DOA, range, Doppler frequency and data symbols jointly. Thus, it can provide optimal results for data decoding and is employed here as a benchmark. In Fig. \ref{MLD}, three cases, which conduct data decoding with the channel estimation results based on $l=3$, $l=5$ and $l=30$ pilots, are considered, while the total number of pilots in frame $v$ is set to $L=30$. In addition, the BS is composed of $N=5$ antennas and the modulation scheme is 8PSK. Due to the fact that the complexity of MLE increases exponentially with the number of estimation parameters, without loss of generality, we consider a single drone 1 in Fig. \ref{MLD}. The result shows that the gap between MRC and MLE in \cite{MLD} reduces with the increase in the number of pilot signals, and MRC can achieve performance comparable to that of MLE when $l$ is large enough. For instance, the gap between MRC and MLE is only $8.2\times 10^{-6}$ when $l=30$ at $\mathrm{SNR} = 9 \;\mathrm{dB}$. This result is due to the fact that the increasing number of pilots has a positive effect in reducing localisation errors, which has been demonstrated in Fig. \ref{pilot number}. The reduction in localisation errors also decreases channel estimation errors because the estimated location parameters are employed to infer channel information, thereby reducing the gap between the data decoding with MRC and MLE. It can also be observed that there is a perfect match between the theoretical and simulated SER of MRC, where $6$th order Taylor expansion in \eqref{Taylor approximation} is considered. 

In Fig. \ref{benchmark}, the performance of the ML-based algorithm is compared to that of CRLB, while the performance of MRC is compared with that of MMSE \cite{MMSE}. Notably, MMSE is employed here as a benchmark with perfect CSI knowledge. In the simulation, $L=5$ pilots are employed to estimate the locations of drone 1 and drone 2 in frame $v$, and 8PSK is adopted as the modulation scheme. As can be observed from Fig. \ref{benchmark}, the performance of the ML-based algorithm approaches that of CRLB across the whole SNR range and even converges with it when SNR is high, which indicates the efficiency of the ML-based localisation algorithm. It can also be found from Fig. \ref{benchmark} that the performance gap between MRC and MMSE reduces with the increase in the number of antennas. This can be attributed to the fact that the estimation errors of location parameters decrease when more antennas are utilised in the BS. Furthermore, the analytical and simulated results for SER with MRC match perfectly during the entire SNR range, where the approximation order is $R=6$.

In Fig. \ref{approximation error}, the approximation error of the Taylor approximation method in Sec. \ref{Approximation} is shown, where $L=5$ pilots are employed to estimate the locations of drone 1 and drone 2 in frame $v$ at $\mathrm{SNR}=3 \; \mathrm{dB}$ and the modulation type is 8PSK. It is worth mentioning that the approximation error is obtained by computing the average of the gaps between the analytical and simulated results of SERs corresponding to the two targets by using MRC. In addition, three cases including $N=5$, $N=7$ and $N=30$ are considered. The result shows that the approximation error reduces quickly with the increase of approximation order $R$ and convergences when $R \geq 5$.

\section{Conclusion}
\label{Conclusion}
This paper presented a comprehensive SER analysis of the PASCAL system, tackling the dual challenges of drone localisation and signal decoding at the BS. An ML-based localisation algorithm was introduced, achieving performance close to the CRLB. Then the estimated location parameters were utilised to infer channel information, enabling effective signal preprocessing via MRC. Using a Taylor approximation method, the average SER was derived in closed form, with simulation results confirming both the rapid convergence and accuracy of the approximation. The findings highlighted a direct relationship between RMSE and SER in the PASCAL system, emphasizing the critical role of precise localisation. Additionally, the analysis showed that increasing the number of pilot signals, although introduced some latency, significantly enhanced both localisation accuracy and communication reliability. The close alignment of analytical and simulation results validated the proposed approach.

\begin{appendices}

\section{The Complete Derivations of ${I}_{5}$ and ${I}_{13}$}
\label{E6 derivation}

To begin, $I_5$ and ${I_{13}}$ can be given by using a general expression as 
 \begin{equation}
    {I_{\psi}} = \int_{\Delta \psi_{\min}}^{\Delta \psi_{\max}} {{\rm{e}}^{ \pm j{C_{13}{\Delta {\psi _k}}{\rm{ - }}\frac{{\rm{1}}}{{{\rm{2}}\sigma _\psi ^2}}{\Delta {\psi _k}}^2}}} d {\Delta {\psi _k}},
\end{equation}
where $C_{13} \in \{C_{11},C_{12} \}$ and $ {\Delta \psi _k} \in \{\Delta \theta_k, \Delta f_{D,k}\}$. ${\Delta \psi_{\min}}$ and ${\Delta \psi_{\max}} $ represents the minimum and maximum values of ${\Delta {\psi _k}}$, and $\sigma _\psi$ indicates the variance of ${\Delta {\psi _k}}$. 

By performing some algebraic operations, ${I_{\psi}}$ can be written as
\begin{equation}
    {I_{\psi}}=\int_{\Delta \psi_{\min}}^{\Delta \psi_{\max}} {{{\rm{e}}^{{{-\left(\sqrt {  \frac{1}{{2\sigma _\psi ^2}}} {\Delta {\psi _k}} \mp \frac{{j{C_{13}}}}{{2\sqrt {  \frac{1}{{2\sigma _\psi ^2}}} }}\right)}^2} - \frac{ C^2_{13}\sigma _\psi ^2}{2}}}} d{\Delta {\psi _k}},
\end{equation}

Afterwards, by using the change-of-variable method in calculating integrals,  ${I_{\psi}}$ can be evaluated as
\begin{equation}
\label{I5 I13}
\begin{array}{l}
    {I_{\psi}}= \displaystyle  \frac{{\sqrt {2\pi \sigma _\psi ^2} {{\rm{e}}^{ - \frac{{C^2_{13}}\sigma _\psi ^2}{2}}}}}{2}\int_{{u_{\min }}}^{{u_{\max }}} {\frac{{{\rm{2}}{{\rm{e}}^{ - {u^2}}}}}{{\sqrt \pi  }}} du      \\  \ \ \ \ \ 
    = \displaystyle  \frac{{\sqrt {2\pi \sigma _\psi ^2} {{\rm{e}}^{ - \frac{{C^2_{13}}\sigma _\psi ^2}{2}}}\left[ {\rm{erf}}(u_{max})-{\rm{erf}}(u_{min}) \right]}}{2} ,\!\!\!\!\!
\end{array}
\end{equation}
where $u=\frac{  {\sqrt 2} \Delta {\psi _k} \mp j{\sqrt 2} C_{13}\sigma^2_\psi }{2 \sigma_\psi}$. As a consequence, $u_{\max}=\frac{  {\sqrt 2} \Delta \psi_{\max} \mp j{\sqrt 2} C_{13}\sigma^2_\psi }{2{\sigma_\psi} }$ and $ u_{\min}=\frac{  {\sqrt 2 {\Delta \psi_{\min}}} \mp j{\sqrt 2 }{C_{13}}\sigma^2_\psi }{2\sigma_\psi }$. 

\end{appendices}

\end{document}